\documentclass[sn-mathphys,Numbered]{sn-jnl}

\usepackage{graphicx}%
\usepackage{multirow}%
\usepackage{amsmath,amssymb,amsfonts}%
\usepackage{amsthm}%
\usepackage{mathrsfs}%
\usepackage[title]{appendix}%
\usepackage{xcolor}%
\usepackage{textcomp}%
\usepackage{manyfoot}%
\usepackage{algorithm}%
\usepackage{algorithmicx}%
\usepackage{algpseudocode}%
\usepackage{listings}%

\usepackage{environ}

\NewEnviron{ShrinkMath}{%
    \begin{equation}
    \scalebox{0.9}{$\BODY$}
    \end{equation}
    }

\usepackage{tcolorbox}
\usepackage{multicol}

\usepackage{xcolor}
\usepackage{mathrsfs}
\usepackage{mathtools}
\usepackage{multirow}
\usepackage{amsmath}
\usepackage{tabularx}
\usepackage{amssymb}
\usepackage{graphicx}

\usepackage{lineno}
\usepackage{caption}
\usepackage{algorithm}
\usepackage{algpseudocode}

\makeatletter
\usepackage{enumitem}

\usepackage{xcolor}
\usepackage{lscape}

\usepackage{savesym}
\numberwithin{equation}{section}
\usepackage{caption}
\usepackage{subcaption}
\usepackage{verbatim}
\usepackage{epsfig}
\usepackage{asymptote}
\graphicspath{{Figures/}} 

\savesymbol{amalg}
\usepackage{mathabx}
\restoresymbol{MAT}{amalg}

\savesymbol{ulcorner}
\savesymbol{urcorner}
\savesymbol{llcorner}
\savesymbol{lrcorner}
\usepackage{amsfonts}
\restoresymbol{AMS}{ulcorner}
\restoresymbol{AMS}{urcorner}
\restoresymbol{AMS}{llcorner}
\restoresymbol{AMS}{lrcorner}
\usepackage[bbgreekl]{mathbbol}
\DeclareSymbolFontAlphabet{\mathbbm}{bbold}
\DeclareSymbolFontAlphabet{\mathbb}{AMSb}

\usepackage{amsthm}
\usepackage{cases}
\usepackage{algorithm}
\usepackage{algpseudocode}
\usepackage{pict2e}
\usepackage{mwe}
\usepackage{footmisc}
\usepackage[cal=boondox,scr=boondoxo]{mathalfa}
\usepackage{multirow}
\usepackage{array}
\usepackage{colortbl}

\usepackage{soul}
\setstcolor{red}
\usepackage{todonotes}

\usepackage{enumitem}

\usepackage{setspace}

\usepackage{url}
\usepackage{mwe}
\usepackage{float}

    \AfterLastShipout{\immediate\write\@mainaux{\relax}}%

\newif\ifHNNdraft
\HNNdrafttrue

\PassOptionsToPackage{normalem}{ulem}
\usepackage{ulem}

\newcommand\footnoteref[1]{\protected@xdef\@thefnmark{\ref{#1}}\@footnotemark}

\renewcommand{\footnotesize}{\scriptsize}

\usepackage{nomencl}
\makenomenclature

\newcommand{\bl}[1]{\boldsymbol{#1}}

\newcommand{\usf}[1]{\boldsymbol{\mathsf #1}}
\newcommand{\busf}[1]{\overline{\boldsymbol{\mathsf #1}}}

\newcommand{\pr}[1]{\left( #1 \right)}

\newcommand{\ag}[1]{\left[ #1 \right]}
\newcommand{\cbr}[1]{\left\{ #1 \right\}}

\newcommand{\set}[2]{\left\{#1\, \big|\, #2\right\}}

\newcommand{\vast}{\bBigg@{4}}
\newcommand{\Vast}{\bBigg@{5}}

\DeclareMathAlphabet{\mathpzc}{OT1}{pzc}{m}{it}

\newcommand{\sprdregion}{\sfkappa_{\rm R}\ag{\mathcal{B}^{\rm sprd}}}

\newcommand{\gelregion}{\sfkappa_{\rm R}\ag{\mathcal{B}^{\rm sub}}}

\definecolor{carmine}{rgb}{0.59, 0.0, 0.09}

\makeatletter
\def\@gobbleappendixname#1\csname thesection\endcsname{\Alph{section}.\arabic{subsection}}
\g@addto@macro{\appendix}{\renewcommand{\p@subsection}{\@gobbleappendixname}}
\makeatother

\newcommand*{\inlineequation}[2][]{%
  \begingroup
    \refstepcounter{equation}%
    \ifx\\#1\\%
    \else
      \label{#1}%
    \fi
    \relpenalty=10000 %
    \binoppenalty=10000 %
    \ensuremath{%
      #2%
    }%
    ~\@eqnnum
  \endgroup
}

\makeatother

\makeatletter
\newcommand*\@dblLabelI {}
\newcommand*\@dblLabelII {}
\newcommand*\@dblequationAux {}

\def\@dblequationAux #1,#2,%
    {\def\@dblLabelI{\label{#1}}\def\@dblLabelII{\label{#2}}}

\newcommand*{\doubleequation}[3][]{%
    \par\vskip\abovedisplayskip\noindent
    \if\relax\detokenize{#1}\relax
       \let\@dblLabelI\@empty
       \let\@dblLabelII\@empty
    \else 
       \@dblequationAux #1,%
    \fi
    \makebox[0.5\linewidth-1.5em]{%
     \hspace{\stretch2}%
     \makebox[0pt]{$\displaystyle #2$}%
     \hspace{\stretch1}%
    }%
    \makebox[0.5\linewidth-1.5em]{%
     \hspace{\stretch1}%
     \makebox[0pt]{$\displaystyle #3$}%
     \hspace{\stretch2}%
    }%
    \makebox[3em][r]{(%
  \refstepcounter{equation}\theequation\@dblLabelI,
  \refstepcounter{equation}\theequation\@dblLabelII)}%
  \par\vskip\belowdisplayskip
}
\makeatother

\newcommand{\inter}{\star}

\usepackage{fonttable}

\usepackage[LGR,T1]{fontenc}
\usepackage{amsmath,etoolbox}
\DeclareSymbolFont{sfletters}{OML}{cmbrm}{m}{it}
\DeclareMathSymbol{\salpha}{\mathord}{sfletters}{"0B}
\DeclareMathSymbol{\sbeta}{\mathord}{sfletters}{"0C}
\DeclareMathSymbol{\sLambda}{\mathord}{sfletters}{'3}

\DeclareMathSymbol{\sgamma}{\mathord}{sfletters}{"0D}

\DeclareMathSymbol{\skappa}{\mathord}{sfletters}{"14}

\DeclareFontFamily{U}{calligra}{}
\DeclareFontShape{U}{calligra}{b}{n}{<->callig15}{}

\DeclareMathSymbol{\sA}{\mathord}{sfletters}{"41}
\DeclareMathSymbol{\sG}{\mathord}{sfletters}{"47}

\newcommand{\sfbitLatinFont}{\usefont{OML}{iwona}{bx}{it}}

\newcommand{\sfmitLatinFont}{\usefont{OML}{iwona}{m}{it}}

\newcommand{\sfmnLatinFont}{\usefont{OML}{iwona}{m}{u}}

\newcommand{\declaresfbLatin}[2]{%
    \protected\csdef{sfbit#1}{\mathord{\text{\sfbitLatinFont#2}}}%
}

\newcommand{\declaresfLatin}[2]{%
    \protected\csdef{sfmit#1}{\mathord{\text{\sfmitLatinFont#2}}}%
}

\newcommand{\declaresfmnLatin}[2]{%
    \protected\csdef{sfmn#1}{\mathord{\text{\sfmnLatinFont#2}}}%
}

\declaresfbLatin{a}{A}
\declaresfbLatin{G}{G}
\declaresfLatin{G}{G}
\declaresfmnLatin{G}{G}

\newcommand{\declaresfgreek}[2]{%
    \protected\csdef{sf#1}{\mathord{\text{\sfgreekfont#2}}}%
}
\newcommand{\sfgreekfont}{\usefont{LGR}{cmss}{m}{it}}
\declaresfgreek{alpha}{a}
\declaresfgreek{beta}{b}
\declaresfgreek{gamma}{g}
\declaresfgreek{kappa}{k}

\declaresfgreek{Lambda}{L}

\newcommand{\declarebsfgreek}[2]{%
    \protected\csdef{bsf#1}{\mathord{\text{\bsfgreekfont#2}}}%
}
\newcommand{\bsfgreekfont}{\usefont{LGR}{cmss}{bx}{it}}
\declarebsfgreek{alpha}{a}
\declarebsfgreek{beta}{b}
\declarebsfgreek{gamma}{g}
\declarebsfgreek{kappa}{k}

\begin{document}
\title[Article Title]{A mechanics theory for the exploration of a high-throughput, sterile 3D \emph{in vitro} traumatic brain injury model}

\author[1]{\fnm{Yang} \sur{Wan}}\email{yang\_wan@brown.edu}
\equalcont{These authors contributed equally to this work.}

\author[2,3]{\fnm{Rafael} \sur{D. Gonz{\'a}lez-Cruz}}\email{rafael\_gonzalez\_cruz@brown.edu}
\equalcont{These authors contributed equally to this work.}

\author[2,3,4]{\fnm{Diane} \sur{Hoffman-Kim }}\email{diane\_hoffman-kim@brown.edu}

\author*[1]{\fnm{Haneesh} \sur{Kesari}}\email{haneesh\_kesari@brown.edu}

\affil[1]{\orgdiv{School of Engineering}, \orgname{Brown University}, \orgaddress{ \city{Providence}, \postcode{02912}, \state{RI}, \country{USA}}}

\affil[2]{\orgdiv{Department of Neuroscience}, \orgname{Brown University}, \city{Providence}, \postcode{02912}, \state{RI}, \country{USA}}

\affil[3]{\orgdiv{Robert J. and Nancy D. Carney Institute for Brain Science}, \orgname{Brown University}, \city{Providence}, \postcode{02906}, \state{RI}, \country{USA}}

\affil[4]{\orgdiv{Center for Biomedical Engineering}, \orgname{Brown University}, \city{Providence}, \postcode{02912}, \state{RI}, \country{USA}}

\abstract{
Brain injuries resulting from mechanical trauma represent an ongoing global public health issue.
Several \textit{in vitro} and \textit{in vivo} models for traumatic brain injury (TBI) continue to be developed for delineating the various complex pathophysiological processes involved in its onset  and progression.
Developing an \textit{in vitro} TBI model that is based on cortical spheroids 
is especially of great interest currently because they 
can replicate key aspects of \textit{in vivo} brain tissue, including its electrophysiology, physicochemical microenvironment, and extracellular matrix composition.
Being able to mechanically deform the spheroids is a key requirement in any effective \textit{in vitro} TBI model.
The spheroids’ shape and size, however, make mechanically loading them, especially in a  high-throughput, sterile, and reproducible manner, quite challenging.
To address this challenge, we present an idea for a spheroid-based, \textit{in vitro} TBI model in which the spheroids are mechanically loaded by being spun by a centrifuge. (An experimental demonstration of this new idea will be published shortly elsewhere.) 
An issue that can limit its utility and scope is that imaging  techniques used in 2D and 3D \textit{in vitro} TBI models  cannot be readily applied in it to determine spheroid strains.
In order to address this issue, we developed a continuum mechanics-based theory to estimate the spheroids’ strains when they are being spun at a constant angular velocity.
The mechanics theory, while applicable here to a special case of the centrifuge-based TBI model, is also of general value since it can help with the further exploration and development of TBI models.}

\keywords{TBI, Brain, Trauma, Mechanobiology, Cell Mechanics, Continuum}

\maketitle

\newpage
\section{Introduction}
\label{sec:Introduction}
Traumatic brain injury (TBI) affects around 55 million people around the world each year and represents an ongoing global public health issue \cite{Maas2022}.
Its prevalence and incidence are higher than other common neurological diseases, including stroke, Alzheimer’s and Parkinson’s diseases \cite{Maas2022}. In the US and Europe, 190--225 patients die every day after suffering from a TBI event and tens of thousands suffer from chronic neurodegenerative diseases and complications resulting from the injury \cite{TBIdata, Maas2022}. The high mortality and  long-term disability associated with TBI highlight the need for further research into its treatment and diagnosis. Currently there are no FDA approved treatments \cite{FDA}  for TBI.  Traumatic brain injury  is often diagnosed using a combination of standard-of-care imaging techniques such as computed tomography (CT) \cite{Mass2005} and magnetic resonance imaging (MRI) scans \cite{Mcdonald2012}, as well as neurological scales assessing consciousness like Glasgow Coma Scale \cite{Teasdale1974}.

Traumatic brain injury is a disease process rather than an event \cite{masel2010traumatic}.
For developing effective treatments, it is critical to understand both TBI’s onset (primary injuries) and its progression (secondary injuries). As such, several \emph{in vivo} as well as \emph{in vitro} TBI models have been developed for delineating the various complex pathophysiological processes involved in its onset and progression.
\emph{In vivo} TBI studies have relied extensively on rodent models. They include injury modalities such as {controlled cortical impact (CCI) \cite{clark1994neutrophil}, fluid percussion injury (FPI) \cite{mcintosh1989traumatic}, weight drop \cite{feeney1981responses}, and sustained focal compression \cite{lin2010ascorbic}. As a whole, \emph{in vivo} injury models are very attractive because they are all-encompassing: they include the brain’s vasculature and structural organization, the brain’s multiple, distinct cell types, the blood-brain barrier, and access to blood circulation and peripheral immune cells. That is, they allow the study of complex multicellular, mechanistic, and systems-level responses to TBI, including axonal demyelination, blood brain barrier breakdown, peripheral immune cell-mediated inflammation, and neurocognitive impairment.

The \emph{in vivo} injury models, however, can also have a few limitations. (1) The \emph{in vivo} models’ all-encompassing nature also makes them difficult to interpret. Specifically, in \emph{in vivo} injury models it is difficult to delineate how the various cellular, biochemical, and biophysical processes affect each other. The various processes have a complex interdependence on each other, and form feedback loops that drive the disease progression. (2) The \emph{in vivo} models can be expensive, and (3) difficult to use.
(4) In \emph{in vivo} models, it is difficult to visualize brain tissue deformation in real time and correlate injury severity to those deformations. Such correlations can potentially provide valuable information for developing inertial-sensor-system-based diagnosis techniques for mild TBI (mTBI) \cite{Rahaman2020,wan2022,wan2023finite,carlsen2021quantitative}.

\emph{In vitro} TBI models in most cases circumvent the limitations (1)-(3) of the \emph{in vivo} models. That is, they are easier to interpret, are less expensive, are easier to use, and pose fewer ethical concerns compared to their \emph{in vivo} counterparts. The earliest \emph{in vitro} studies involved subjecting neuronal and glial monolayers (2D cell cultures) to higher pressures and monitoring plasma membrane damage and cell death \cite{Murphy1993}. In fact, the 2D \emph{in vitro} models do not even suffer from limitation (4) of the \emph{in vivo} models, since it is straightforward to monitor deformations in them using a time sequence of microscopy images and image processing algorithms. Despite their many attractive features, 2D \emph{in vitro} models suffer from one major limitation: they may not be sophisticated enough to capture the primary pathophysiological processes involved in TBI.


Three dimensional (3D) cell culture models are a relatively recent development \cite{Hanna2023}.
They are generally more sophisticated than 2D cell culture models.
In particular,  cortical spheroids---which are a special type of 3D cell culture models---replicate key aspects of \emph{in vivo} brain tissue, such as the electrophysiology, dimensionality, physicochemical microenvironment, and the extracellular matrix composition observed \cite{dingle2015three, shoemaker2021biofidelic} \emph{in vivo}. Hence, currently there is significant interest in developing \emph{in vitro} TBI models that are based on 3D cell culture models.
Some recent works in this direction involve compressing neuronal cells embedded in hydrogels made of specific extracelullar matrix proteins or biomaterial composites \cite{bar2016strain, liaudanskaya2020modeling}, and brain organoids \cite{shoemaker2021biofidelic}. The two main challenges in developing 3D  \emph{in vitro}  TBI models are (1) designing a mechanical loading system that can be used to mimic the mechanical forces that \emph{in vivo} tissue experiences during TBI while maintaining high throughput, and sterility in the experiment, and (2) being able to estimate the deformations experienced by the \emph{in vitro} tissue during the experiment.

We propose a new 3D \emph{in vitro} TBI model in which mechanical loads are applied to cortical spheroids, via the aid of centrifugal forces.
Cortical spheroids are grown within a 3D soft substrate, i.e., within the cavities (such as microwells) molded into the surface of a soft material (see inset in Fig.~\ref{fig:centrifuge} (a)), such as an agarose hydrogel.
Cell culture media bathes the cortical spheroids as well as their soft substrate.
We propose to load the cortical spheroids by spinning them along with their soft substrates and their fluid media using a centrifuge (see Fig.~\ref{fig:centrifuge} (a)).
The centrifuge's angular velocity can vary during the experiment.
In a frame that rotates with the centrifuge’s rotating arm, the cortical spheroid, the soft substrate, and the fluid media experience body forces that  push them away from the axis of rotation (Fig.~\ref{fig:centrifuge} (b)).
These forces cause the fluid media and the soft substrate to push the cortical spheroid’s surfaces that they are respectively in contact with towards each other; thus squeezing the spheroid (Fig.~\ref{fig:centrifuge} (c)).
We will experimentally demonstrate in a follow up publication that this centrifugation based method of loading the cortical spheroids can provide the basis for the development of a high-throughput and sterile \emph{in vitro} 3D TBI model.

\begin{figure}[H]
    \centering
\includegraphics[width=\textwidth]{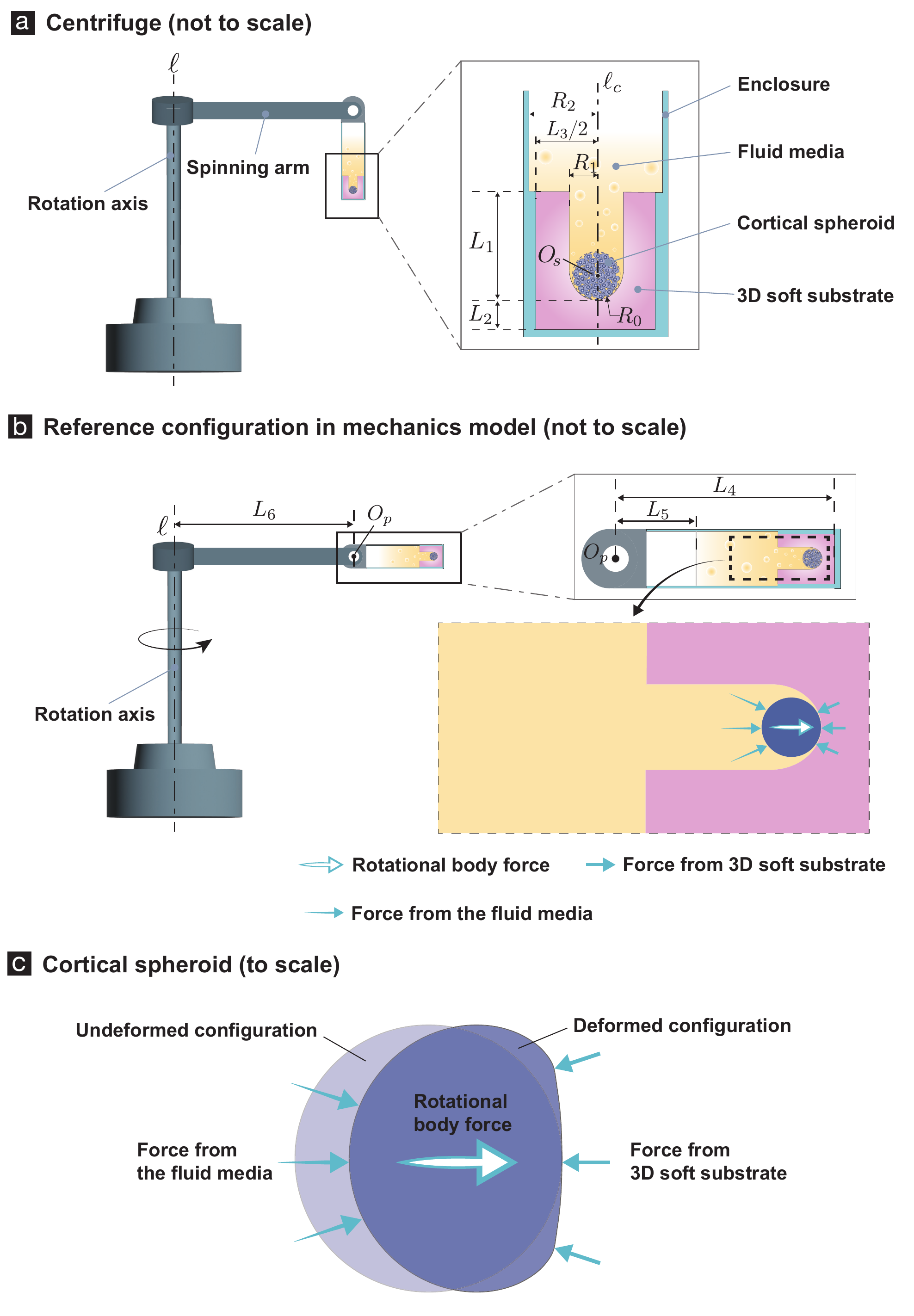}
    \caption{
    An illustration of the centrifuge-TBI-model. For ease of exposition, especially when we develop the mechanics theory in \S\ref{sec:MM}, we only show one cortical spheroid as being spun in (a) and (b). However, in practice several thousands of spheroids can be spun simultaneously.}
    \label{fig:centrifuge}
\end{figure}

The above proposed centrifugation based 3D \emph{in vitro} TBI model (centrifuge-TBI-model) does not have the limitations (1)--(3) of the \emph{in vivo} models, neither does it have the limitation of insufficient sophistication of the 2D \emph{in vitro} models.
The model is designed to be  high throughput; maintain conditions that are as sterile as those present at the time the spheroids are grown; and allow loading of the spheroids in a reproducible manner.

 The centrifuge-TBI-model is of high throughput, since thousands of spheroids can be grown simultaneously, and then spun simultaneously as well\footnote{In Fig.~\ref{fig:centrifuge} we only show a single cortical spheroid being spun, even though in practice several thousand can be spun simultaneously. We do so since showing a single spheroid makes it easier to use Fig.~\ref{fig:centrifuge} to explain the development of the mechanics theory.}; with no steps between the two that involve processing the spheroids serially, such as individual pipetting, positioning, or probing. These features are possible for the following reasons. The 3D soft substrate is micromachined to have close to a hundred cavities/microwells on it. A single spheroid (which could be as large as 8000 cells) grows in a cavity via self-assembly given the nonadhesive properties of the soft substrate. It has already been demonstrated that using the 3D soft substrates several thousands of spheroids can be grown simultaneously \cite{dingle2015three}.   A single substrate is usually smaller than a cubic centimeter in size, therefore a lab grade centrifuge will be able to hold over a hundred substrates. Hence, several thousands of spheroids can be spun simultaneously if desired.
In addition to the growth and loading operations in the centrifuge-TBI-model each being parallel in nature, another aspect of the model that critically contributes to it being high throughput is that the spheroids are loaded \textit{in situ}, i.e., they are tested in the same substrate as that in which they are grown (compare, e.g., Fig.~\ref{fig:centrifuge} (a) and (b)).

The  loading conditions are sterile due to the \textit{in situ} testing, and because the loading on the spheroids is being primarily performed by the same fluid media and the soft substrate used to grow them (e.g., see Figs.~\ref{fig:centrifuge} (c) and (d)).

The mechanical loading, i.e., the force on the spheroids, in the experiment can be easily and robustly tuned via the centrifuge's angular velocity and the volume of  the fluid media.


However, as with all models, the proposed model too has some limitations. One of the most significant of those is the same as limitation (4) of the \emph{in vivo} models, which, to reiterate, is the inability to visualize tissue deformation in real time and correlate injury severity to those deformations. In order to address this limitation, in this paper we consider a special case of the centrifuge-TBI-model and develop a mechanics theory for determining the cortical spheroids' deformation in it. The special case we consider is the one which the centrifuge's angular velocity is constant as a function of time. The primary assumptions in our mechanics theory are described in \S\ref{sec:assumps}. The mathematical preliminaries necessary for detailing our theory are described in \S\ref{sec:mathpreli}. We develop the theory in  \S\ref{sec:MM}, and summarize it in \S\ref{sec:CBVPs}. Results from numerical solutions of our theory for two representative values of  centrifuge angular velocity  are presented in \S\ref{sec:Results}.

The proposed centrifuge-TBI-model, in theory, has many advantages compared to other models.
The theory developed in this paper  applies to a special case of the centrifuge-TBI-model, and it provides an indirect means of determining the deformations.
Future studies will work toward a general and direct approach for determining deformations.
We believe that our theory is of value since it allows us to explore the centrifuge-TBI-model and its viability and potential for \textit{in vitro} TBI studies.


\section{Primary assumptions and modeling decisions, and their underlying rationale}
\label{sec:assumps}

In this section we list some of the primary assumptions and modeling decisions that we made for developing our theory to estimate the deformations in the cortical spheroid as it is spun by the centrifuge.

As we already mentioned in \S\ref{sec:Introduction}, we restrict ourselves to the case where the centrifuge is being operated at a constant angular velocity of $\omega_{\rm max}~{\rm rad/s}$. We made this decision in order to reduce the theory's complexity.

For modeling the deformations of the cortical spheroid, we  also model the motion and deformations of the 3D soft substrate containing it, and the fluid media that bathes the spheroid and the substrate.

 We assume that the deformations and stresses in the spheroid and the substrate's region that is in its vicinity are axi-symmetric.

We assume that the mechanics  of the spheroid, the substrate, and the fluid media can be well modeled using continuum theories~\cite{Gurtin1982}.
Consequently, we model the spheroid and the substrate as homogeneous solids, and the fluid media as a homogeneous fluid. Even more specifically, we model the spheroid as a spherical ball composed of an incompressible neo-Hookean material, the substrate as composed of a compressible neo-Hookean material, and  the fluid media as an incompressible Newtonian fluid. The constitutive law for the fluid media is given in \S\ref{sec:PCM} and those for the spheroid and substrate are given in \S\ref{sec:govequ}.
Additionally, we model the interaction between the spheroid and substrate as non-adhesive frictionless contact.
We refer to  \textit{(i)} the cortical spheroid, \textit{(ii)} the 3D soft substrate, and \textit{(iii)} the fluid media collectively as the continua.




We ignore  acceleration due to gravity in our theory.

We assume that in a frame that rotates with the centrifuge's spinning arm, all mechanical fields remain stationary w.r.t. time. (see \S\ref{sec:concluding} for a discussion of this important assumption).


On account of the previous assumption it follows that the  motion of the continua can be described as
\begin{equation}
\bl{x}_{\boldsymbol{\tau}}[\bl{X}]=\bl{Q}_{\bl{\tau}}\bl{I}_{\mathbb{E}_{\rm R}\to \mathbb{E}}\pr{\bl{X}+\bl{U}^{\inter}\ag{\bl{X}}}.
\label{eq:motion}
\end{equation}
Here $\bl{X}$ is the position vector of a continuum material particle $\mathcal{X}$ in a reference configuration. (The particle can belong to the spheroid, the soft substrate, or the fluid media.)
We will henceforth be referring to the material particle $\mathcal{X}$ by its reference position vector, $\bl{X}$.
We call $\bl{x}_{\bl{\tau}}\ag{\cdot}$ the deformation map and $\bl{x}_{\bl{\tau}}\ag{\bl{X}}$ the material particle $\bl{X}$'s current position vector at the time instance $\bl{\tau}$.
We call $\bl{U}^{\inter}\ag{\bl{X}}$ the intermediate displacements of the material particle $\bl{X}$.
In general the intermediate displacements in addition to $\bl{X}$ also depend on the time instance $\bl{\tau}$ . In our theory the intermediate displacements only depending on $\bl{X}$ too is a consequence of our previous assumption.
We define the symbols $\bl{Q}_{\boldsymbol{\tau}}$, and $\bl{I}_{\mathbb{E}_{\rm R}\to \mathbb{E}}$ appearing in \eqref{eq:motion} in \S\ref{sec:DeformationMapping}.
For a mathematically complete and rigorous formulation of \eqref{eq:motion} see  \cite{wan2023finite}.

The material field $\bl{U}^{\inter}\ag{\cdot}$ is an unknown \textit{a priori}. The strains and the stresses in the continua depend on the values of its gradient, $\left\{\bl{\nabla}_{\bl{X}}\ag{\bl{U}^{\star}}\right\}\ag{\cdot}$.
In \S\ref{sec:MM} we derive the equations whose solution (cf. \S\ref{sec:CBVPs} and \S\ref{sec:Results}) will yield $\bl{U}^{\inter}\ag{\cdot}$, and hence the strains and the stresses.


\begin{figure}[H]
    \centering
        \includegraphics[width=\textwidth]{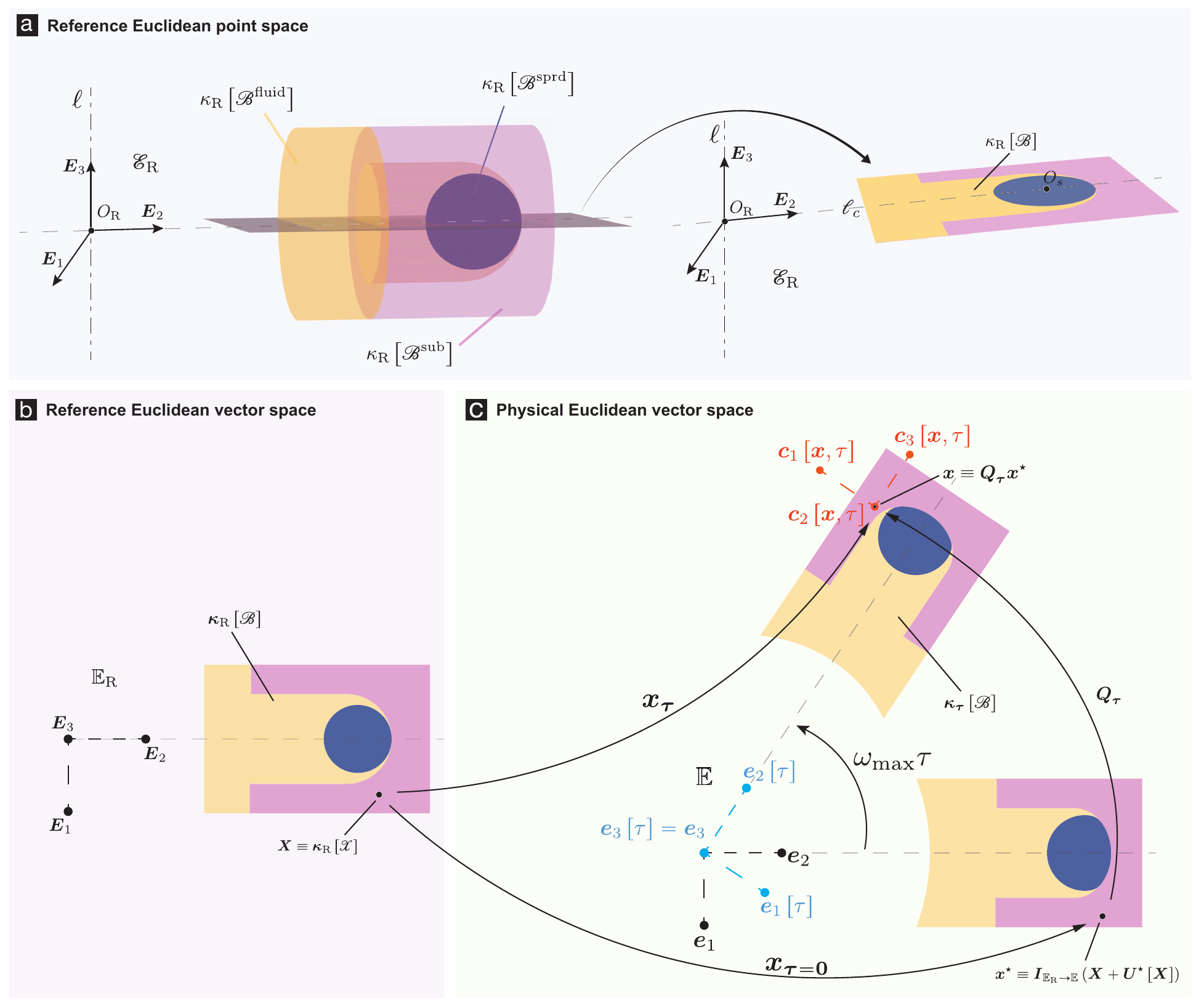}
    \caption{
    An illustration of the various mathematical objects that we use in our mechanics model of spheroid centrifugation. All objects are defined in \S\ref{sec:mathpreli} and \S\ref{sec:KinematicsinFRSS2DElasticModel}.
  }
    \label{fig:notion}
\end{figure}

\section{Mathematical preliminaries}
\label{sec:mathpreli}
In this section we  present the preliminary  mechanics and mathematical notions that are needed for the development of our theory. Some of these notions can also be found in  \cite[\S2.1]{wan2022} and \cite[\S2.1]{Rahaman2020}
.

\subsection{Abstract vector spaces in our model}
\label{sec:GeometryinFRSS2DElasticModel}
Let $\mathbb{E}_{\rm R}$ be an oriented Euclidean vector space, i.e., an oriented finite dimensional, real, inner product space, and let the affine point space $\mathcal{E}_{\rm R}$ have $\mathbb{E}_{\rm R}$ as its associated vector translation space.  We refer to $\mathbb{E}_{\rm R}$ and $\mathcal{E}_{\rm R}$ as the reference Euclidean vector and point space, respectively. Let $\mathbb{E}$ and $\mathcal{E}$ be another pair of Euclidean vector and affine point space, respectively.
Our continua  (which can either be the spheroid, the 3D soft substrate, or the fluid media)  execute their motion in $\mathcal{E}$.  For that reason, we refer to $\mathbb{E}$ and $\mathcal{E}$ as the physical Euclidean vector space and point space, respectively. We model each of our continuum bodies, spheroid, soft substrate, and the media using the topological spaces $\mathcal{B}^{\rm sprd}$, $\mathcal{B}^{\rm sub}$, and $\mathcal{B}^{\rm fluid}$, respectively  (see Fig.~\ref{fig:notion} (a)).

We call a select continuous, injective map from $\mathcal{B}$
(which can be $\mathcal{B}^{\rm fluid}$, $\mathcal{B}^{\rm sprd}$, or $\mathcal{B}^{\rm sub}$)
into $\mathbb{E}_{\rm{R}}$ the reference configuration and denoted it as $\boldsymbol{\kappa}_{\rm R}$.
The elements of $\mathcal{B}$ are called material particles.
We call $\boldsymbol{X}\equiv\boldsymbol{\kappa}_{\rm R}\ag{\mathcal{X}}$ the particle $\mathcal{X}$'s reference position vector and $\boldsymbol{\kappa}_{\rm R}\ag{\mathcal{B}}$ the reference body (see Fig.~\ref{fig:notion} (b)).
Taking some arbitrary point $O_{\rm R}\in\mathcal{E}_{\rm R}$ to be $\mathcal{E}_{\rm R}$'s origin (see Fig.~\ref{fig:notion} (a)), to $\bl{\kappa}_{\rm R}$ we associate the map $\kappa_{\rm R}: \mathcal{B} \rightarrow \mathcal{E}_{\rm R}$ such that $O_{\rm R}+\bl{\kappa}_{\rm R}\ag{\mathcal{X}}=\kappa_{\rm R}\ag{\mathcal{X}}$. We call $X\equiv\kappa_{\rm R}\ag{\mathcal{X}}$ the particle $\mathcal{X}$'s reference point.


We model time as a one-dimensional normed vector space $\mathbb{T}$ and denote a typical element in it as $\bl{\tau}=\tau\bl{s}$, where $\tau\in\mathbb{R}$ and $\bl{s}$ is a fixed vector which has units of seconds.


\subsection{Cartesian basis vectors}
The sets $\pr{\bl{E}_{i}}_{i\in\mathcal{I}}$ and $\pr{\bl{e}_{i}}_{i\in\mathcal{I}}$, where $\mathcal{I}:=\pr{1,2,3}$, are orthonormal sets of basis vectors for $\mathbb{E}_{{\rm R}}$ and $\mathbb{E}$, respectively.
By orthonormal we mean that the inner product between $\bl{E}_i$ and $\bl{E}_j$,
or $\bl{e}_{i}$ and $\bl{e}_j$, where $i, j\in\mathcal{I}$, equals $\delta_{ij}$,
the Kronecker delta symbol, which equals unity iff $i=j$ and zero otherwise.
In our problem, we take $\bl{E}_{i}$ and $\bl{e}_{i}$, $i\in\mathcal{I}$, to have the units of meters.
The Cartesian co-ordinates of $X$ which we denote as $\breve{\usf{X}}\ag{X}=\pr{\breve{\usf{X}}_i\ag{X}}_{i\in\mathcal{I}}$, are components of $\bl{X}$ w.r.t. $\bl{E}_i$, that is $\breve{\usf{X}}_i\ag{X}=X_i$, where $X_i:=\bl{X}\cdot\bl{E}_i$.
For simplicity, ${\usf X} \equiv \pr{X_1,X_2,X_3}$.

We denote the space of all $m \times n$ real nested ordered sets, where $m, n\in \mathbb{N}$, as $\mathcal{M}_{m \times n}(\mathbb{R})$.
Thus $\breve{\usf{X}}\ag{X}\in\mathcal{M}_{3 \times 1}(\mathbb{R})$.
We call the map $\mathcal{E}_{\rm{R}}$ $ \ni X \mapsto \breve{\usf{X}}\ag{X}\in \mathcal{M}_{3 \times 1}(\mathbb{R})$ the Cartesian co-ordinate map.
Let $\pr{\usf{x}_i}_{i\in\mathcal{I}}$ be orthonormal sets of basis vectors for $\mathcal{M}_{3 \times 1}(\mathbb{R})$, or $\mathbb{R}^3$, where $\usf{x}_1:=\pr{1,0,0}$, $\usf{x}_2:=\pr{0,1,0}$, and $\usf{x}_3:=\pr{0,0,1}$.
When we refer to $\bl{X} \in \mathbb{E}_{\rm R}$, $\usf{X} \in \mathcal{M}_{3 \times 1}(\mathbb{R})$, or $X \in \mathcal{E}_{\rm R}$ as a material particle we in fact mean the
material particle $\mathcal{X}\in\mathcal{B}$.

\subsection{Co-rotational Cartesian basis vectors for $\mathbb{E}$}

Let
\begin{equation}
\pr{Q_{ij}\ag{\tau}}_{i,j\in \mathcal{I}}
=
\left(\begin{array}{ccc}
\cos\ag{ \omega_{\rm max}\tau}& -\sin\ag{ \omega_{\rm max}\tau} & 0 \\ [3 pt]
 \sin\ag{ \omega_{\rm max}\tau} & \cos\ag{ \omega_{\rm max}\tau} & 0 \\ [3 pt]
 0 & 0 & 1
\end{array}
\right):=:\usf{Q}\ag{\tau}.
\label{eq:QDef}
\end{equation}
The matrix $\usf{Q}\ag{\tau}$ belongs to the special orthonormal group $ SO(3)\subset \mathcal{M}_{3\times 3}\pr{\mathbb{R}}$, and  therefore
satisfies the equations
\begin{subequations}
\begin{align}
\usf{Q}^{\sf T}\ag{\tau}\,\usf{Q}\ag{\tau}&=\usf{I}_{3\times3},\label{equ:QQ:1}
\intertext{and}
\usf{Q}\ag{\tau}\,\usf{Q}^{\sf T}\ag{\tau}&=\usf{I}_{3\times3},\label{equ:QQ:2}
\end{align}
\label{eq:QQ}
\end{subequations}
where $\usf{Q}^{\sf T}\ag{\tau}$ is the transpose of $\usf{Q}\ag{\tau}$, i.e., $\usf{Q}^{\sf T}\ag{\tau}=\pr{\usf{Q}\ag{\tau}}^{\sf T}$, and $\usf{I}_{3\times3}=\pr{\delta_{ij}}_{i,j\in\mathcal{I}} \in \mathcal{M}_{3 \times 3}(\mathbb{R})$.

Using $\usf{Q}\ag{\tau}$ we define the co-rotational set of basis vectors for $\mathbb{E}$, $(\bl{e}_i\ag{\tau})_{i\in \mathcal{I}}$ , as
\begin{equation}
\bl{e}_i\ag{\tau}=Q_{ji}\ag{\tau}\bl{e}_j.
\label{eq:corotationalbasis}
\end{equation}
Note that $(\bl{e}_i\ag{\tau})_{i\in \mathcal{I}}$ change with time (see Fig.~\ref{fig:notion} (c)). However at each time instance $\bl{\tau}$, they form an orthonormal set of  vectors and   provide a basis for $\mathbb{E}$.

\subsection{Co-rotational Cartesian co-ordinates}


Given $\bl{x}\in \mathbb{E}$, let
\begin{equation}
\breve{x}_{i}\ag{\bl{x},\tau}:=\bl{x}\cdot \bl{e}_i\ag{\tau},
\end{equation}
where $\pr{\bl{e}_i\ag{\tau}}_{i\in \mathcal{I}}$ is defined via \eqref{eq:corotationalbasis}.
We call $\pr{\breve{x}_i\ag{\bl{x},\tau}}_{i\in \mathcal{I}}=:\breve{\usf{x}}\ag{\bl{x},\tau}$ the co-rotational Cartesian co-ordinates of $\bl{x}$ at the time instance $\tau$.

\subsection{Co-rotational cylindrical basis vectors for $\mathbb{E}$}

Using $\breve{\usf{x}}\ag{\bl{x},\tau}$ we define the set of co-rotational cylindrical basis vectors $\pr{\bl{c}_i\ag{ \bl{x},\tau}}_{i\in \mathcal{I}}$ (see Fig.~\ref{fig:notion} (c)) at the point $\bl{x}$ at the time instance $\tau$ as
\begin{equation}
\pr{\bl{c}_i\ag{\bl{x},\tau}}_{i\in \mathcal{I}}=
\usf{R}\ag{
\breve{\usf{x}}\ag{\bl{x},\tau}}\pr{\bl{e}_{i}\ag{\tau}}_{i\in \mathcal{I}},
\end{equation}
where $\usf{R}\ag{\cdot}:\mathcal{M}_{3\times 1}\pr{\mathbb{R}}\to \mathcal{M}_{3\times 3}\pr{\mathbb{R}}$,
\begin{equation}
\usf{R}\ag{\pr{x_1,x_2,x_3}}:=\frac{1}{\sqrt{x_1^2+x_3^2}}\left(\begin{array}{ccc}
x_1& 0 & x_3 \\ [3 pt]
 x_3 & 0 & -x_1 \\ [3 pt]
 0 &  \sqrt{x_1^2+x_3^2} & 0
\end{array}
\right).
\end{equation}

\subsection{Linear maps between vector spaces}

Say $\mathbb{W}$ and $\mathbb{U}$ are two arbitrary, oriented Euclidean vector spaces; for instance, they can
be $\mathbb{E}_{\rm R}$ and $\mathbb{E}$. We denote the space of all
linear maps (transformations/operators) from $\mathbb{W}$ to $\mathbb{U}$
as $\ensuremath{\mathcal{L}}(\mathbb{W},\mathbb{U})$.
We denote the norm of a vector $\bl{w}_{1}$ in $\mathbb{W}$ that
is induced by $\mathbb{W}$'s inner product, i.e., $(\bl{w}_{1}\cdot\bl{w}_{1})^{1/2}$, as $\lVert\bl{w}_{1}\rVert$. For $\bl{u}_{1}\in\mathbb{U}$, the expression $\bl{u}_{1}\otimes\bl{w}_{1}$
denotes the linear map from $\mathbb{W}$ to $\mathbb{U}$
defined as
\begin{equation}
\pr{\bl{u}_{1}\otimes\bl{w}_{1}}\bl{w}_{2}=\bl{u}_{1}\pr{\bl{w}_{1}\cdot\bl{w}_{2}},
\end{equation}
where $\bl{w}_{2}\in\mathbb{W}$. If the sets $\pr{\bl{u}_i}_{i\in\mathcal{I}}$
and $\pr{\bl{w}_i}_{i\in\mathcal{I}}$ provide bases for $\mathbb{U}$
and $\mathbb{W}$, respectively, then it can be shown that $\pr{\pr{\bl{u}_{i}\otimes\bl{w}_j}_{j\in\mathcal{I}}}_{i\in\mathcal{I}}$,
which we will henceforth abbreviate as $\pr{\bl{u}_{i}\otimes\bl{w}_j}_{i,j\in\mathcal{I}}$, provides a basis for $\ensuremath{\mathcal{L}}(\mathbb{W},\mathbb{U})$. The number $T_{ij}$, where $i,j\in\mathcal{I}$, is called the component of $\bl{T}\in\mathcal{L}(\mathbb{W},\mathbb{U})$
w.r.t. $\bl{u}_{i}\otimes\bl{w}_j$ iff $T_{ij}=\bl{u}_{i}\cdot\pr{\bl{T}\bl{w}_j}$. We call the nested ordered set $\pr{T_{ij}}_{i,j\in\mathcal{I}}$ the component
form of $\bl{T}$ w.r.t. $\pr{\bl{u}_{i}\otimes\bl{w}_j}_{i,j\in\mathcal{I}}$,
and denote it as $\usf{T}$.

From here on, unless otherwise specified, we will be following the
Einstein summation convention. As per this convention a repeated
index in a term will imply a sum over that term with the repeated index taking values in $\mathcal{I}$. For example, the expression $X_{i}\bl{E}_{i}$ represents the sum $\sum_{i\in\mathcal{I}}X_{i}\bl{E}_{i}$.
And an unrepeated index in a term will signify a set of 3 terms. For example, the term $\bl{E}_{i}$ represents the set $\set{\bl{E}_{i}}{i\in\mathcal{I}}$.

 The operator $D_j\ag{\cdot}$ is defined such that
\begin{equation}
    \{D_j\ag{f}\}\ag{\usf{X}}=\frac{\partial f\ag{\usf{X}}}{\partial X_j},
    \label{equ:Dj}
\end{equation}
for $f:\mathbb{R}^3 \rightarrow \mathbb{R}$. We abbreviate $D_j\ag{f}$ as $D_j f$.

\section{Mechanics Model}
\label{sec:MM}
\subsection{Kinematics}
\label{sec:KinematicsinFRSS2DElasticModel}



\subsubsection{Stationary and reference configurations}

We show our assumed geometries for the continua in the centrifuge-TBI-model in Fig.~\ref{fig:centrifuge} (a) and (b). The configuration shown in (a) is for when the centrifuge is stationary (recall that we have ignored acceleration due to gravity), and the one shown in (b) is the reference configuration in our problem.

In Fig.~\ref{fig:centrifuge} (a),  the spheroid lies in a 3D soft substrate; while the
 3D soft substrate itself lies in an enclosure  connected to the centrifuge's spinning arm.
The geometries of the spheroid, the substrate, and the enclosure are all axi-symmetric about the axis $\ell_{c}$ shown in Fig.~\ref{fig:centrifuge} (a).
Thus, the spheroid is, well, a spherical ball of radius $R_0~\rm m$.
The cavity in the substrate that the spheroid lies in has the shape of a test tube. It is open at the top; it is
$L_1~\rm{m}$ deep; the radius of   its circular cross-sections is $R_1~\rm{m}$; and its base has a hemispherical shape.
The substrate does not completely fill the enclosure. The substrate's cross-sections towards the top are annular disks, while those towards the bottom are circular disks.
The inner radii of the annular disks is $R_1~\rm m$.
The outer radii of the annular disks (and the radii of the circular disks) is $L_3/2~\rm m$.
The enclosure's cross-sections towards the top as well as those towards the bottom are both annular disks, albeit of different inner radii.
The inner radii of the enclosure's annular disks that lie towards the bottom are the same as the outer radii of the substrate's annular disks, namely, $L_3/2~\rm m$.
The inner radii of the enclosure's annular disks that lie towards the top are $R_2~\rm m$.

The spheroid's center $O_s$ lies on the central axis $\ell_c$ (see Fig.~\ref{fig:centrifuge} (a)), and it rests at the  bottom of the cavity, with a single point touching (see Fig.~\ref{fig:centrifuge} (a)). The thickness of the substrate under the spheroid is $L_2~\rm{m}$.

We refer to the point at which the centrifuge’s spinning arm attaches to the enclosure as $O_P$ (marked in Fig.~\ref{fig:centrifuge} (b)). We take the reference configuration for our problem to be the one shown in Fig.~\ref{fig:centrifuge} (b), which is the same as that in Fig.~\ref{fig:centrifuge}
 (a) except that the enclosure and the continua have undergone a rigid body rotation about the axis that is perpendicular to the plane spanned by the centrifuge's rotation axis and spinning arm (shown in Fig.~\ref{fig:centrifuge} (a)) and passing through $O_P$\footnote{Note that in our problem the reference configuration  and the configuration when the centrifuge is not spinning are not isomorphic. When the centrifuge is not spinning the central axis and the rotation axis are parallel to each other, where as in the reference configuration the central axis $\ell_c$ and the rotation axis are perpendicular to each other.}.

The base of the substrate is at a distance of $L_4~\rm{m}$ from the point $O_P$ (shown in Fig.~\ref{fig:centrifuge} (b)).
The spheroid and the substrate are bathed in the fluid media.
The surface of the fluid media is at a distance of $L_5~\rm{m}$ from the point $O_P$ (shown in Fig.~\ref{fig:centrifuge} (b)).
The length of the spinning arm is $L_6~\rm{m}$.

Typical values for all the geometry parameters in the stationary and reference configurations, which are partially based on the measurements reported in \cite{dingle2015three}, are given in Table~\ref{tb:TypicalGeoParams}.

\definecolor{tablecolor1}{RGB}{219,221,223}
\definecolor{tablecolor2}{RGB}{160,183,206}
\setlength{\tabcolsep}{8pt}
\setlength{\arrayrulewidth}{0.2mm}
\begin{table}[h!]
\renewcommand*{\arraystretch}{3}
\centering
\caption{Typical values for the geometry parameters $R_0$--$R_2$, and $L_1$--$L_6$ in the centrifuge-TBI-model.
These parameters are defined in Fig.~\ref{fig:centrifuge} (a)--(b).
These typical values are based on the spheroid, and 3D soft substrate dimensions reported in \cite{dingle2015three}; geometry of the enclosure (marked in Fig.~\ref{fig:centrifuge} (a)) in which the 3D soft substrate containing the spheroid is typically held; the amount of fluid media that is typically added to the enclosure, which is around $1~{\rm ml}$ \cite{dingle2015three}; and the dimensions of a typical lab grade centrifuge, such as 5810R Eppendorf \cite{Centrifuge}. The units of all parameters are \emph{meters}.}
\begin{tabular}{c|ccccccccc}
 \hline
 \rowcolor{tablecolor2}
Parameter name & $R_0 $ & $R_1$ & $R_2$ & $L_1$ & $L_2$&  $L_3$& $L_4$ &$L_5$&$ L_6$\\
 Parameter value $\times 10^{3}$ &
 $0.08 $ & $0.2 $ & $7.96$&  $0.8$&  $1.5$ &  $0.8$& $66$ & $59.61$ & $112.69$\\
\hline
\end{tabular}
\label{tb:TypicalGeoParams}
\end{table}

\subsubsection{Deformation mapping}
\label{sec:DeformationMapping}
The motion of the continua is given by \eqref{eq:motion}.
To partially reiterate,
the vector $\bl{X}\in \mathbb{E}_{\rm R}$ is the reference position vector of the material particle $\mathcal{X}$.
The symbol
$\bl{I}_{\mathbb{E}_{\rm R}\rightarrow \mathbb{E}}$ denote the identity linear map from $\mathbb{E}_{\rm R}$ onto $\mathbb{E}$.
More explicitly, $\bl{I}_{\mathbb{E}_{\rm R}\rightarrow \mathbb{E}}=\bl{e}_{i}\otimes\bl{E}_{i}$.
Without loss of generality, we take that the continua rotate about $\bl{e}_3$.
Since we restrict ourselves to  the case in which the continua rotate at a fixed angular velocity of $\omega_{\rm max}\ \rm rad/s$, the assumption of rotation  about $\bl{e}_3$ implies that $\bl{Q}_{\bl{\tau}}=Q_{ij}\ag{\tau}\bl{e}_{i}\otimes\bl{e}_{j}$, where $Q_{ij}\ag{\tau}$ are defined in \eqref{eq:QDef}.
The map $\bl{U}^{\inter}:\mathbb{E}_{\rm R} \rightarrow \mathbb{E}_{\rm R}$ is the intermediate displacement field of $\mathcal{B}$.
The symbol $\bl{x}_{\bl{\tau}}\ag{\bl{X}}$ is the material particle $\bl{X}$'s position vector in $\mathbb{E}$ at the time instance $\bl{\tau}$.
The set $\bl{\kappa}_{\bl{\tau}}\ag{\mathcal{B}}=\set{\bl{x}_{\bl{\tau}}\ag{\bl{X}}\in\mathbb{E}}{\bl{X}\in\bl{\kappa}_{\rm R}\ag{\mathcal{B}}}$ is called the current body (see Fig.~\ref{fig:notion} (c)).

As per \eqref{eq:motion} the continua's deformations are time invariant in the co-rotational basis $(\bl{e}_i\ag{\tau})_{i\in \mathcal{I}}$. The co-rotational basis themselves rotate about the time stationary  vector $\bl{e}_3$ with the constant angular velocity $\omega_{\rm max}~{\rm rad/s}$ (see Fig.~\ref{fig:notion} (c)). 



\subsubsection{Displacements components}

Expressing $\bl{X}=X_i\bl{E}_i$, and $\bl{U}^{\inter}\ag{\bl{X}}=U_{i}^{\inter}\ag{\usf{X}}\bl{E}_i$, and using \eqref{eq:motion} and \eqref{eq:corotationalbasis} it can be shown that
\begin{subequations}    \begin{equation}
\bl{x}_{\tau}\ag{\bl{X}}=\overline{x}_i\ag{\usf{X}}\bl{e}_i\ag{\tau},\end{equation} where
\begin{align}
    \overline{x}_{i}\ag{\usf{X}}&=X_i+U^{\inter}_i\ag{\usf{X}}.
\label{eq:DispComponents}
\end{align}
\label{eq:DeformationMapping}
\end{subequations}
Denoting $(\overline{x}_{i}\ag{\usf{X}})_{i\in \mathcal{I}}$ as $\busf{x}\ag{\usf{X}}$ and $\pr{U^{\inter}_i\ag{\usf{X}}}_{i\in \mathcal{I}}$ as $\usf{U}^{\inter}\ag{\usf{X}}$, \eqref{eq:DispComponents} can equivalently be expressed as
\begin{equation}
\busf{x}\ag{\usf{X}}=\usf{X}+\usf{U}^{\inter}\ag{\usf{X}}.
\label{eq:DispND}
\end{equation}

Let
\begin{align}
\sfkappa_{\rm R} \ag{\mathcal{B}^{\rm sprd}}&:=\left\{\usf{X}\in \mathbb{R}^3~|~X_i \bl{E}_i\in \bl{\kappa}_{\rm R}\ag{\mathcal{B}^{\rm sprd}}\right\},
\label{eq:sprdDomain}
\intertext{and}
\sfkappa_{\rm R} \ag{\mathcal{B}^{\rm sub}}&:=\left\{\usf{X}\in \mathbb{R}^3~|~X_i \bl{E}_i\in \bl{\kappa}_{\rm R}\ag{\mathcal{B}^{\rm sub}}\right\}.
\label{eq:gelDomain}
\end{align}
We refer to the restriction of $\busf{x}\ag{\cdot}$ to $\sfkappa_{\rm R} \ag{\mathcal{B}^{\rm sprd}}$ as $\busf{x}^{\rm sprd}\ag{\cdot}$. The maps $\busf{x}^{\rm sub}\ag{\cdot}$ and $\busf{x}^{\rm fluid}\ag{\cdot}$ are defined similarly.



\subsubsection{Velocity components}
We call $\mathcal{L}\pr{\mathbb{T},\mathbb{E}}$ the physical velocity
vector space and denote it as $\mathbb{V}$.
It can be shown that the
set $\pr{\bl{v}_{i}\ag{\tau}}_{i\in\mathcal{I}}$, where $\bl{v}_{i}\ag{\tau}\in\mathbb{V}$
and are defined such that $\{\bl{v}_{i}\ag{\tau}\}\bl{\tau}=\tau\bl{e}_i\ag{\tau}$, that is $\bl{v}_{i}\ag{\tau}:=\bl{e}_i\ag{\tau}\otimes\bl{s}^{*}$, where $\bl{s}^{*}$ is the dual of $\boldsymbol{s}$, provides
an orthonormal basis for $\mathbb{V}$. The velocity of a material
particle $\bl{X}$ executing its motion in $\mathbb{E}$ lies in $\mathbb{V}$.
The velocity of the material particle $\bl{X}$ at the instant $\bl{\tau}$, which we denote as $\bl{V}_{\bl{\tau}}\ag{\bl{X}}$, equals the value of the Fr\'echet derivative of the map $\mathbb{T}\ni\bl{\tau}\mapsto\bl{x}_{\bl{X}}\ag{\bl{\tau}}\in\mathbb{E}$,
where $\bl{x}_{\bl{X}}\ag{\bl{\tau}}=\bl{x}_{\bl{\tau}}\ag{\bl{X}}$, at the time instance $\bl{\tau}$.
Thus, it follows from \eqref{eq:motion} that for $\tau \ge 0$
\begin{subequations}
\begin{align}
\bl{V}_{\boldsymbol{\tau}}\ag{\bl{X}}&=V_{i}\ag{\usf{X}}\bl{v}_i\ag{\tau},
\label{eq:defvel}
\end{align}
where
\begin{align}
V_{i}\ag{\usf{X}}&=W_{ij}\pr{X_j+U^{\inter}_{j}\ag{\usf{X}}},
\label{eq:VelComponents}
\intertext{and}
W_{ij}&=Q'_{kj}[\tau]Q_{ki}\ag{\tau}.
\label{eq:WComponents}
\end{align}
\end{subequations}
From \eqref{eq:QDef} and \eqref{eq:WComponents} it follows that
\begin{equation}
    \pr{W_{ij}}_{i,j\in \mathcal{I}}=\left(
\begin{array}{ccc}
 0 & -\omega_{\rm max} & 0 \\
 \omega_{\rm max} & 0 & 0 \\
 0 & 0 & 0 \\
\end{array}
\right)=:\usf{W}.
\label{eq:WComps}
\end{equation}
Denoting $(V_{i}\ag{\usf{X}})_{i\in \mathcal{I}}$ as $\usf{V}\ag{\usf{X}}$, \eqref{eq:VelComponents} can equivalently be written as
\begin{equation}
\usf{V}\ag{\usf{X}}=\usf{W}\pr{\usf{X}+\usf{U}^{\inter}\ag{\usf{X}}}.
\end{equation}

The velocity of the material particle located at $\bl{x} \in \mathbb{E}$ at the time instance $\bl{\tau}$ is defined as
\begin{equation}
\bl{v}_{\bl{\tau}}\ag{\bl{x}}=\bl{V}_{\tau}\ag{\bl{x}_{\bl{\tau}}^{-1}\ag{\bl{x}}}.
\label{equ:vel2}
\end{equation}
From \eqref{eq:defvel}, \eqref{eq:VelComponents}, and \eqref{eq:DispComponents}, the equation \eqref{equ:vel2} can be written as
\begin{equation}
\bl{v}_{\bl{\tau}}\ag{\bl{x}}=W_{ij}\breve{x}_j\ag{\bl{x},\tau}\bl{v}_i\ag{\tau}.
\label{equ:vel3}
\end{equation}

\subsubsection{Accelerations}

We call $\mathcal{L}\pr{\mathbb{T},\mathbb{V}}$ the physical acceleration
vector space and denote it as $\mathbb{A}$. It can be shown that the
set $\pr{\bl{a}_{i}\ag{\tau}}_{i\in\mathcal{I}}$, where $\bl{a}_{i}\ag{\tau}\in\mathbb{A}$
and are defined such that $\{\bl{a}_{i}\ag{\tau}\}\bl{\tau}=\tau\bl{v}_i\ag{\tau}$, i.e., $\bl{a}_i\ag{\tau}=\bl{v}_i\ag{\tau}\otimes\bl{s}^{*}$, provides
an orthonormal basis for $\mathbb{A}$. The acceleration of a material
particle $\bl{X}$ executing its motion in $\mathbb{E}$ lies in $\mathbb{A}$.
The acceleration of $\bl{X}$ at the time instance $\bl{\tau}$
equals the value of the Fr\'echet derivative of the map $\mathbb{T}\ni\bl{\tau}\mapsto\bl{V}_{\bl{X}}(\bl{\tau})\in\mathbb{V}$, where $\bl{V}_{\bl{X}}(\bl{\tau})=\bl{V}_{\bl{\tau}}(\bl{X})$, at the
time instance $\bl{\tau}$. Thus, it follows from \eqref{eq:defvel}  that for $\tau \ge 0$
\begin{subequations}
\begin{align}
\bl{A}_{\tau}\ag{\bl{X}}&=A_{i}\ag{\usf{X}}\bl{a}_i\ag{\tau},
\label{equ:acce}
\intertext{where}
A_{i}\ag{\usf{X}}&=W_{im}W_{mp}\pr{X_p+U^{\inter}_{p}\ag{\usf{X}}}.
\label{eq:AccelComponents}
\end{align}\end{subequations}

Denoting $(A_{i}\ag{\usf{X}})_{i\in \mathcal{I}}$ as $\usf{A}\ag{\usf{X}}$, \eqref{eq:AccelComponents} can be equivalently be written as
\begin{equation}
\usf{A}\ag{\usf{X}}=\usf{W}^2\pr{\usf{X}+\usf{U}^{\inter}\ag{\usf{X}}}.
\end{equation}

The acceleration of the material particle located at $\bl{x} \in \mathbb{E}$ at the time instance $\bl{\tau}$ is defined as
\begin{equation}
\bl{a}_{\bl{\tau}}\ag{\bl{x}}=\bl{A}_{\tau}\ag{\bl{x}_{\bl{\tau}}^{-1}\ag{\bl{x}}}.
\label{equ:acce2}
\end{equation}
From \eqref{equ:acce}, \eqref{eq:AccelComponents}, and \eqref{eq:DispComponents}, the equation \eqref{equ:acce2} can be written as
\begin{equation}
\bl{a}_{\bl{\tau}}\ag{\bl{x}}=W_{im}W_{mp}\,\breve{x}_{p}\ag{\bl{x},\tau}\bl{a}_i\ag{\tau}.
\label{equ:acce3}
\end{equation}

\subsubsection{Deformation gradient and Strains}
The deformation gradient corresponding to the deformation mapping $\bl{x}_{\bl{\tau}}\ag{\cdot}$, given in \eqref{eq:motion}, is
\begin{subequations}
\begin{align}
\{\bl{\nabla}_{\bl{X}}\ag{\bl{x}_{\bl{\tau}}}\}\ag{\bl{X}}&=:\bl{F}_{\boldsymbol{\tau}}\ag{\bl{X}}=F_{ij}\ag{\usf{X}}\bl{e}_{i}\ag{\tau}\otimes \bl{E}_j,
\intertext{where}
F_{ij}\ag{\usf{X}}&:=\delta_{ij}+D_{j}U^{\inter}_{i}\ag{\usf{X}}.
\label{eq:DGCompos}
\end{align}
\end{subequations}
The right Cauchy-Green deformation tensor corresponding to the deformation gradient $F_{ij}\ag{\usf{X}}\bl{e}_i\ag{\tau}\otimes \bl{E}_j$ is
\begin{subequations}
\begin{equation}
\bl{C}\ag{\bl{X}}=C_{ij}\ag{\usf{X}}\bl{E}_{i}\otimes \bl{E}_j,
\end{equation}
where
\begin{equation}
C_{ij}\ag{\usf{X}}=F_{mi}\ag{\usf{X}}F_{mj}\ag{\usf{X}}.
\end{equation}
\end{subequations}
We abbreviate $\pr{F_{ij}\ag{\usf{X}}}_{i,j\in \mathcal{I}}$ and $\pr{C_{ij}\ag{\usf{X}}}_{i,j\in \mathcal{I}}$ as $\usf{F}\ag{\usf{X}}$ and $\usf{C}\ag{\usf{X}}$, respectively.

\subsection{Equilibrium}

\subsubsection{Cauchy-momentum equations in the reference body}
\label{sec:balanceequref}
 It follows from the principle of balance of linear momentum and our  decision to ignore acceleration due to gravity that
\begin{equation}
\{\text{\rm Div}\, \bl{F}_{\boldsymbol{\tau}}\bl{S}\}\ag{\bl{X}}
=\rho_{\rm o}\bl{A}_{\tau}\ag{\bl{X}},
\label{eq:equilibrium}
\end{equation}
 where $\{\text{\rm Div}\, \bl{F}_{\bl{\tau}}\bl{S}\}\ag{\cdot}$ is the divergence of the field $\bl{X}\mapsto \bl{F}_{\bl{\tau}}\ag{\bl{X}}\bl{S}\ag{\bl{X}}$.
 Here $\bl{S}\ag{\bl{X}}$ is the $2^{\rm nd}$ Piola-Kirchhoff stress tensor at the material particle $\bl{X}$.

In component form \eqref{eq:equilibrium} can be written as
\begin{equation}    \{D_j\ag{F_{im}S_{mj}}\}\ag{\usf{X}}=\rho_0 A_{i}\ag{\usf{X}},
\label{eq:EquilComponentForm}
\end{equation}
where $S_{ij}\ag{\usf{X}}$, $i$,$j\in \mathcal{I}$, are the components of $\bl{S}\ag{\bl{X}}$\footnote{\label{TSij}Here we omit providing the mathematical details of how precisely $S_{ij}\ag{\usf{X}}$ and $T_{ij}\ag{\usf{x}}$ are, respectively, related to $\bl{S}\ag{\bl{X}}$ and $\bl{T}_{\boldsymbol{\bl{\tau}}}\ag{\bl{x}}$. Since doing so will require notions from \emph{exterior algebra} and \emph{differential geometry} that need a significant amount of space to properly explain, and hence would distract from the primary focus of this paper.  }.

In \eqref{eq:EquilComponentForm} replacing $A_{i}\ag{\usf{X}}$ with the RHS of \eqref{eq:AccelComponents} we get
\begin{align}
\{D_j\ag{F_{im}S_{mj}}\}\ag{\usf{X}}&=\rho_0 W_{im}W_{mp}\pr{X_p+U^{\star}_{p}\ag{\usf{X}}},
\label{eq:equil3}
\end{align}
where $\rho_0~{\rm kg/m^3}$ is the density of the continua.

Noting from \eqref{eq:WComps} that $W_{im}W_{mp}=-\omega_{\rm max}^2\pr{\delta_{ip}-\delta_{i3}\delta_{3p}}$ in \eqref{eq:equil3} we get
\begin{align}
\{D_j\ag{F_{im}S_{mj}}\}\ag{\usf{X}}&=-\rho_0 \omega_{\rm max}^2 \pr{X_i+U^{\star}_{i}\ag{\usf{X}}-\delta_{i3}\pr{X_3+U^{\star}_{3}\ag{\usf{X}}}}.
\label{eq:equil4}
\end{align}

The domain of \eqref{eq:equil4} is either $\sfkappa_{\rm R} \ag{\mathcal{B}^{\rm sprd}}$, or $\sfkappa_{\rm R} \ag{\mathcal{B}^{\rm sub}}$, which were, respectively, defined in  \eqref{eq:sprdDomain} and \eqref{eq:gelDomain}. Irrespective, of whether $\usf{X}$ belongs to $\sfkappa_{\rm R} \ag{\mathcal{B}^{\rm sprd}}$ or $\sfkappa_{\rm R} \ag{\mathcal{B}^{\rm sub}}$ the co-ordinates $X_1$ and $X_3$ are always less than $L_3/2$, where recall that $L_3/2$ is the outer radii of the substrate, and is $0.4\times10^{-3}$ in our model.
The co-ordinate $X_2$ in the domains, however, varies between $176\times10^{-3}$ and $179\times10^{-3}$. (Fig.~\ref{fig:centrifuge} can help in understanding how we arrived at these ranges for the different co-ordinates.)    Therefore, in \eqref{eq:equil4} we ignore $X_1$, and $X_3$,  in comparison to $X_2$.
The intermediate displacement components $U^{\inter}_i\ag{\usf{X}}$ are unlikely to be larger than the height of the substrate, which is, roughly, $3~{\rm millimeters}$.
Therefore, we also ignore $U_i^*\ag{\usf{X}}$ in comparison to $X_2$ in \eqref{eq:equil4}. In summary, we approximate \eqref{eq:equil4} as



\begin{equation}
\{D_j\ag{F_{im}S_{mj}}\}\ag{\usf{X}}=-\rho_0 \omega_{\rm max}^2 \delta_{i2}\delta_{2j}X_j.
\label{eq:equilibrium4}
\end{equation}

\subsubsection{Cauchy-momentum equations in the current body}
It follows from the principle of balance of linear momentum  that
\begin{equation}
\{\text{\rm Div}\, \bl{T}_{\boldsymbol{\tau}}\}\ag{\bl{x}}
=\rho_{\tau}\ag{\bl{x}}\bl{a}_{\bl{\tau}}\ag{\bl{x}},
\label{eq:equilibriumcurrent}
\end{equation}
for all $\bl{x}\in \bl{\kappa}_{\bl{\tau}}\ag{\mathcal{B}}$, where $\{\text{\rm Div}\, \bl{T}_{\bl{\tau}}\}\ag{\cdot}$ is the divergence  of the field $\bl{x}\mapsto \bl{T}_{\bl{\tau}}\ag{\bl{x}}$.
 Here $\bl{T}_{\boldsymbol{\bl{\tau}}}\ag{\bl{x}}$ is the Cauchy stress tensor at the current position $\bl{x}$ at the time instance $\bl{\tau}$. And $\rho_{\tau}\ag{\bl{x}}:=\rho_0/\textsf{Det}\ag{\bl{F}_{\bl{\tau}}\ag{\bl{x}_{\bl{\tau}}^{-1}\ag{\bl{x}}}}$, where $\textsf{Det}\ag{\cdot}$ is the determinant operator.


In \eqref{eq:equilibriumcurrent} replacing $\bl{a}_{\bl{\tau}}\ag{\bl{x}}$ with the RHS of \eqref{equ:acce3} and rewriting \eqref{eq:equilibriumcurrent} in component form, we get
\begin{align}
\{D_jT_{ij}\}\ag{\usf{x}}&=\overline{\rho}_{\tau}\ag{\usf{x}}W_{im}W_{mp}x_p,
\label{eq:equilibriumcurrent1}
\end{align}
for all $\usf{x}\in \sfkappa_{\tau}\ag{\mathcal{B}}$,
\begin{equation}
\sfkappa_{\tau}\ag{\mathcal{B}}:=\left\{\usf{x}:=\pr{x_1,x_2,x_3}\in \mathbb{R}^3~|~x_i \bl{e}_i\ag{\tau}\in \bl{\kappa}_{\bl{\tau}}\ag{\mathcal{B}}\right\},
\end{equation}
and $T_{ij}\ag{\usf{x}}$, $i$,$j\in \mathcal{I}$, are the components of $\bl{T}_{\boldsymbol{\bl{\tau}}}\ag{\bl{x}}$\footnotemark[2].
Here $\overline{\rho}_{\tau}\ag{\cdot}$ is defined such that $\overline{\rho}_{\tau}\ag{\usf{x}}=\rho_{\tau}\ag{x_i\bl{e}_i\ag{\tau}}$.
Noting from \eqref{eq:WComps} that $W_{im}W_{mp}=-\omega_{\rm max}^2\pr{\delta_{ip}-\delta_{i3}\delta_{3p}}$ in \eqref{eq:equilibriumcurrent1} we get
\begin{align}
\{D_jT_{ij}\}\ag{\usf{x}}&=-\overline{\rho}_{\tau}\ag{\usf{x}}\omega_{\rm max}^2 \pr{x_i-\delta_{i3}x_3}.
\label{eq:equilibriumcurrent2}
\end{align}




\subsection{Pressure in the fluid media}
\label{sec:PCM}
We model the fluid media as an incompressible Newtonian fluid.
It can be shown that in our problem, the rate of deformation tensor is naught (see \S\ref{sec:D0} for details).
Consequently, from \cite[\S22]{Gurtin1982},
we can get
\begin{equation}
    \usf{T}\ag{\usf{x}}=-p^{\textrm{fluid}}_{{s}}\ag{\usf{x}}\usf{I}_{3\times3},
    \label{eq:pressure}
\end{equation}
for all $\usf{x}\in \sfkappa_{\tau}\ag{\mathcal{B}^{\rm fluid}}$,
\begin{equation}
\sfkappa_{\tau}\ag{\mathcal{B}^{\rm fluid}}:=\left\{\usf{x}\in \mathbb{R}^3~|~x_i \bl{e}_i\ag{\tau}\in \bl{\kappa}_{\bl{\tau}}\ag{\mathcal{B}^{\rm fluid}}\right\},
\end{equation}
and $\usf{T}\ag{\usf{x}}:=\pr{T_{ij}\ag{\usf{x}}}_{i,j\in \mathcal{I}}$ and $p^{\rm fluid}_{s}\ag{\cdot}$ is the pressure field.

In \eqref{eq:equilibriumcurrent2} substituting $\overline{\rho}_{\tau}\ag{\usf{x}}$ as $\rho_0$, since we have assumed the fluid media as being incompressible, and $\usf{T}[\usf{x}]$ as $-p^{\rm fluid}_{s}\ag{\usf{x}}\usf{I}_{3\times3}$ from \eqref{eq:pressure},
we get that
\begin{equation}
\left\{D_i\ag{p^{\rm fluid}_{s}}\right\}\ag{\usf{x}}=\rho_0\omega_{\rm max}^2 \pr{x_i-\delta_{i3}x_3}, \quad \forall \usf{x}\in \sfkappa_{\tau}\ag{\mathcal{B}^{\rm fluid}}.
\label{equ:governingpressure1}
\end{equation}
It can be shown that in our problem the free surface of the fluid at the time instance $\bl{\tau}$ (marked as $\partial\usf{B}^{\rm fluid}_{\tau,5}$ in Fig.~\ref{fig:boundarynotion} (a)) is always  part of a cylinder. More specifically, it can be shown that
\begin{equation}
\partial\usf{B}^{\rm fluid}_{\tau,5}=\{\usf{x}\in \sfkappa_{\tau}\ag{\mathcal{B}^{\rm fluid}}~|~x_1^2+x_2^2=l_7^2\}.
\label{eq:GammaTau5Def}
\end{equation}
The parameter $l_7$ in \eqref{eq:GammaTau5Def} is the distance of the center of $\partial\usf{B}^{\rm fluid}_{\tau,5}$ from the rotation axis (see Fig.~\ref{fig:boundarynotion} (a)).

The surface $\partial\usf{B}^{\rm fluid}_{\tau,5}$ experiences the atmospheric  pressure $p^{\rm atm}\ \rm Pa$, where $p^{\rm atm}=1.01325\times  10^5 $. Hence, one of the boundary conditions on $p^{\rm fluid}_{s}\ag{\cdot}$ is
\begin{equation}
p^{\rm fluid}_{s}\ag{\usf{x}}=p^{\rm atm},\quad \usf{x}\in \partial\usf{B}^{\rm fluid}_{\tau,5},
\label{equ:conpressure1}
\end{equation}

Solving \eqref{equ:governingpressure1} with the boundary condition \eqref{equ:conpressure1} we get that
\begin{equation}
p^{\rm fluid}_{s}\ag{\usf{x}}=\frac{1}{2}\rho_0\omega_{\rm max}^2\pr{x_1^2+x_2^2-l_7^2}+p^{\rm atm}.
\label{equ:pressure}
\end{equation}



\begin{figure}[t]
    \centering
        \includegraphics[width=\textwidth]{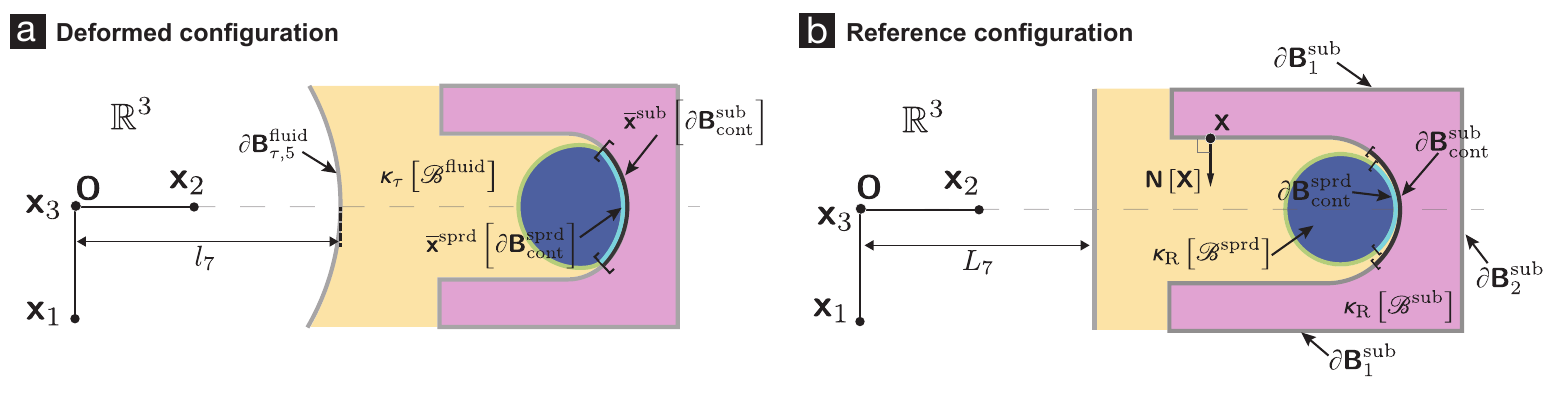}
    \caption{
    An illustration of  the cortical spheroid, the fluid media, and the soft substrate surfaces in the deformed (a) and reference (b) configurations.}
    \label{fig:boundarynotion}
\end{figure}



\subsection{Material constitutive laws}
\label{sec:govequ}

\subsubsection{Constitutive law for the spheroid}
In \eqref{eq:equilibrium4} when $\usf{X}
\in \sfkappa_{\rm R} \ag{\mathcal{B}^{\rm sprd}}$

\begin{subequations}
\begin{equation}
S_{ij}\ag{\usf{X}}=\breve{S}_{ij}^{\rm sprd}\ag{\usf{C}\ag{\usf{X}}}-p^{\rm sprd}_{m}\ag{\usf{X}} J\ag{\usf{C}\ag{\usf{X}}} \left(\usf{C}\ag{\usf{X}}\right)^{-1},
\label{eq:SpCE}
\end{equation}
where $S_{ij}\ag{\usf{X}}$ is the $i$-$j^{\rm th}$ component of $\bl{S}\ag{\bl{X}}$, the $2^{\rm nd}$ Piola Kirchhoff stress at the material particle $\bl{X}$,
\begin{equation}
    J\ag{\usf{C}}:=\sqrt{\textsf{Det}\ag{\usf{C}}},
\end{equation}
$\breve{S}_{ij}^{\rm sprd}\ag{\cdot}$ is the $i$-$j^{\rm th}$ component of $\breve{\usf{S}}^{\rm sprd}\ag{\cdot}$,
\begin{equation}
    \breve{\usf{S}}^{\rm sprd}[\usf{C}]=\mu\, \textsf{Det}\ag{\rm \usf{C}}^{-\frac{1}{3}}\pr{\usf{I}_{3\times3} - \frac{1}{3}\textsf{ Tr}\ag{\usf{C}}\usf{C}^{-1}},
\label{eq:ShearContEq}
\end{equation}
\label{eq:CLSpheroid}
\end{subequations}
and $\mu~{\rm Pa}$ is the shear modulus.
Here  $\textsf{Tr}\ag{\cdot}$ is the trace operator.
In \eqref{eq:SpCE} the quantity $p^{\rm sprd}_{m}\ag{\usf{X}}$ acts as a Lagrange undetermined multiplier, which can be interpreted as the hydrostatic pressure at the material particle $\usf{X}$.
Since $p^{\rm sprd}_{m}\ag{\cdot}$ is an unknown \textit{a priori}, we solve \eqref{eq:equilibrium4} in conjunction with the incompressibility constraint
\begin{equation}
J\ag{\usf{C}\ag{\usf{X}}}=1.
\label{equ:incompressibility}
\end{equation}

Equations~\eqref{eq:CLSpheroid} are the incompressible neo-Hookean material model from \cite[equation 5.50]{bonet1997}.

\subsubsection{Constitutive law for the 3D soft substrate}
In \eqref{eq:equilibrium4} when $\usf{X}\in  \sfkappa_{\rm R} \ag{\mathcal{B}^{\rm sub}}$
\begin{subequations}
    \begin{equation}
    S_{ij}[\usf{X}]=\breve{S}_{ij}^{\rm sub}\ag{\usf{C}\ag{\usf{X}}},
    \label{eq:Sgel}
\end{equation}
where $\breve{S}_{ij}^{\rm sub}\ag{\cdot}$ is the $i$-$j^{\rm th}$ component of $\breve{\usf{S}}^{\rm sub}\ag{\cdot}$,
\begin{equation}
    \breve{\usf{S}}^{\rm sub}\ag{\usf{C}}=\mu\, \textsf{Det}\ag{\rm \usf{C}}^{-1/3}\pr{\usf{I}_{3\times3} - \frac{1}{3}\textsf{ Tr}\ag{\usf{C}}\usf{C}^{-1}}+\pr{\lambda+\frac{2}{3}\mu}J\ag{\usf{C}}\pr{J\ag{\usf{C}}-1}\usf{C}^{-1}.
\end{equation}
\label{eq:CLGel}
\end{subequations}
Here $\lambda~{\rm Pa}$ and $\mu~{\rm Pa}$ are the Lam\'e parameters.
Equations~\eqref{eq:CLGel} are the compressible neo-Hookean material model from \cite[\S3.5.5]{bower2009applied}.

\subsection{Boundary conditions}
\label{sec:bc}

The solution of \eqref{eq:equilibrium4} also requires the use of the following boundary conditions.

Let $\partial \usf{B}^{\rm sprd}$ and $\partial \usf{B}^{\rm sub}$
be the surfaces of the spheroid $\sfkappa_{\rm R} \ag{\mathcal{B}^{\rm sprd}}$ and the substrate $\sfkappa_{\rm R} \ag{\mathcal{B}^{\rm sub}}$, respectively.
Let $\partial \usf{B}^{\rm sprd}_{\rm cont}$ and $\partial \usf{B}^{\rm sub}_{\rm cont}$ be the surfaces of $\sfkappa_{\rm R} \ag{\mathcal{B}^{\rm sprd}}$ and $\sfkappa_{\rm R} \ag{\mathcal{B}^{\rm sub}}$, respectively, that come into contact with each other (see Fig.~\ref{fig:boundarynotion} (b)). They are both unknown \textit{a priori}.
The boundary conditions on $\partial \usf{B}^{\rm sprd}_{\rm cont}$ and $\partial \usf{B}^{\rm sub}_{\rm cont}$ are that there are no shear tractions on them, and the displacements of the spheroid and substrate on them respectively are such that
\begin{equation}
\busf{x}^{\rm sprd}\ag{\partial\usf{B}^{\rm sprd}_{\rm cont}}= \busf{x}^{\rm sub}\ag{\partial\usf{B}^{\rm sub}_{\rm cont}},
\label{eq:ContactBC}
\end{equation}
see Fig.~\ref{fig:boundarynotion} (a) for an illustration.

Let $\partial \usf{B}^{\rm sub}_1$ and $\partial \usf{B}^{\rm sub}_2$ be the surfaces of $\sfkappa_{\rm R} \ag{\mathcal{B}^{\rm sub}}$ shown in Fig.~\ref{fig:boundarynotion} (b).
\begin{subequations}
    The boundary conditions on $\partial \usf{B}^{\rm sprd}\setminus \partial \usf{B}^{\rm sprd}_{\rm cont}$ and $\partial \usf{B}^{\rm sub}\setminus \partial \usf{B}^{\rm sub}_{\rm cont}\setminus \bigcup_{i=1}^{2} \partial \usf{B}^{\rm sub}_{i}$ (see Fig.~\ref{fig:boundarynotion} (b)), due to the spheroid's and substrate's, respective, interactions with the fluid media are
\begin{equation}
    \pr{\usf{C}\ag{\usf{X}}\usf{S}\ag{\usf{X}}-p^{\rm fluid}_{m}\ag{\usf{X}}\usf{I}_{3\times3}}\usf{N}\ag{\usf{X}}= \usf{0}_{3\times 1},
\label{eq:SpBC1}
\end{equation}
where $\usf{S}\ag{\usf{X}}:=\pr{S_{ij}\ag{\usf{X}}}_{i,j\in \mathcal{I}}$, $\usf{0}_{3\times 1}\equiv\pr{0,0,0}$, and $\usf{N}\ag{\usf{X}}$ is the unit outward surface normal vector at the location $\usf{X}$ (see Fig.~\ref{fig:boundarynotion} (b) for example).
The field $p^{\rm fluid}_{m}\ag{\usf{X}}$ in \eqref{eq:SpBC1} is
$
p^{\rm fluid}_{s}\ag{\busf{x}\ag{\usf{X}}}$, where $p^{\rm fluid}_{s}\ag{\cdot}$ is given in \eqref{equ:pressure}.
More concretely,
\begin{equation}
p^{\rm fluid}_{m}\ag{\usf{X}}=
\frac{1}{2}\rho_0\omega_{\rm max}^2\pr{\pr{X_1+U_1^{\star}\ag{\usf{X}}}^2+\pr{X_2+U_2^{\star}\ag{\usf{X}}}^2-l_7^2}+p^{\rm atm}.
\label{eq:pflMat}
\end{equation}
We cannot independently calculate $l_7$ in our model. Therefore, in \eqref{eq:pflMat} we approximate $l_7$ as $L_7$, the distance of the fluid's free surface to the rotation axis under the assumption that none of the continua (the spheroid, the fluid, and the substrate) deform (see Fig.~\ref{fig:boundarynotion} (b)). Also, since $U_1^{\star}$, $X_1$, and $U_2^{\star}$ are much smaller than $X_2$, in \eqref{eq:pflMat} we approximate $\pr{X_1+U_1^{\star}\ag{\usf{X}}}$ as naught, and $\pr{X_2+U_2^{\star}\ag{\usf{X}}}$ as $X_2$. In summary, we compute
\begin{equation}
p^{\rm fluid}_{m}\ag{\usf{X}}\approx
\frac{1}{2}\rho_0\omega_{\rm max}^2\pr{X_2^2-L_7^2}+p^{\rm atm}.
\label{eq:pflMat1}
\end{equation}

\end{subequations}

An additional boundary condition on the substrate is that
\begin{subequations}
    \begin{align}
U^{\star}_1\ag{\usf{X}}X_1&=-U^{\star}_3\ag{\usf{X}}X_3, \quad \forall \usf{X}\in \partial\usf{B}^{\rm sub}_1\label{eq:agabvp2}    \\
U^{\star}_2\ag{\usf{X}}&=0, \quad \forall \usf{X}\in \partial\usf{B}^{\rm sub}_2.
\label{eq:agabvp3}
\end{align}
\end{subequations}
The boundary condition \eqref{eq:agabvp2} is a consequence of setting the radial component of the displacement field on  $\partial\usf{B}^{\rm sub}_1$ (see Fig.~\ref{fig:boundarynotion} (b)) to be naught, which we do to model the constraint from the enclosure (see Fig~.\ref{fig:centrifuge} (a)).
We choose the boundary condition \eqref{eq:agabvp3} to model the fact that the substrate sits in the  enclosure (see Fig~.\ref{fig:centrifuge} (a)), which constrains its deformation in the $\bl{E}_2$ direction on $\partial\usf{B}^{\rm sub}_2$.

\section{Coupled boundary value problems}
\label{sec:CBVPs}
As per our model, the motion of the spheroid and the 3D soft substrate is given by the family of deformation maps $\bl{x}_{\bl{\tau}}\ag{\cdot}$.
This family of deformation maps can be constructed using   \eqref{eq:DeformationMapping} once the displacement field components $U_{i}^{\star}\ag{\cdot}$ are known.
The restrictions of $U_{i}^{\star}\ag{\cdot}$ to $\sfkappa_{\rm R}\ag{\mathcal{B}^{\rm sprd}}$ (resp. $\sfkappa_{\rm R}\ag{\mathcal{B}^{\rm sub}}$ ) are obtained by solving the partial differential equation (PDE) \eqref{eq:equilibrium4} over the region $\sfkappa_{\rm R}\ag{\mathcal{B}^{\rm sprd}}$ (resp. $\sfkappa_{\rm R}\ag{\mathcal{B}^{\rm sub}}$).
We refer to the PDE \eqref{eq:equilibrium4} posed over the region $\sfkappa_{\rm R}\ag{\mathcal{B}^{\rm sprd}}$ as the spheroid boundary value problem (BVP), and the PDE \eqref{eq:equilibrium4} posed over the region $\sfkappa_{\rm R}\ag{\mathcal{B}^{\rm sub}}$ as the substrate BVP.
Recall that the functions $\left\{F_{im}S_{mj}\right\}\ag{\cdot}$ appearing in   \eqref{eq:equilibrium4} are defined as
\begin{equation}
\usf{X}\mapsto F_{im}\ag{\usf{X}}S_{mj}\ag{\usf{X}}
\label{eq:FuncFS}.
\end{equation}

In the spheroid (resp. substrate) BVP the $F_{im}\ag{\cdot}$ in \eqref{eq:FuncFS} are to be interpreted as the restrictions of the $F_{im}\ag{\cdot}$ defined in \eqref{eq:DGCompos} to $\sprdregion$ (resp. $\gelregion$).
For the spheroid (resp. substrate) BVP the $S_{ij}\ag{\cdot}$ in \eqref{eq:FuncFS}  is given by the function \eqref{eq:CLSpheroid} (resp. \eqref{eq:CLGel}). In the spheroid BVP, due to the presence of the Lagrange multiplier (pressure field) $p^{\rm sprd}_{m}\ag{\cdot}$ in \eqref{eq:CLSpheroid}, the PDE \eqref{eq:equilibrium4} needs to be  solved jointly with the incompressibility constraint equation \eqref{equ:incompressibility}. The boundary conditions in the spheroid and the substrate BVPs are detailed in \S\ref{sec:bc}.

Note that the contact boundary condition \eqref{eq:ContactBC} is part of both the spheroid and the substrate BVPs. It couples the two BVPs, since it involves displacement components from both BVPs. Therefore, the two BVPs cannot be solved independently. We solve the spheroid and the substrate BVPs simultaneously using finite element techniques.

\section{Representative numerical solutions of the theory}
\label{sec:Results}
To get a preliminary understanding into the type of strains and stresses that the cortical spheroids experience in the constant angular velocity centrifuge-TBI-model, and for demonstrating our theory, we compute various strain and stress measures in the cortical spheroids using our theory for some representative values of angular velocity, geometry parameters, and material properties.

Typical lab grade centrifuges are capable of reaching top speeds in the range of 1000--5000 revolution per minute (RPM). Therefore, for the representative calculations we consider angular velocities of $2000$ RPM, i.e., 209 radians per second (${\rm rad/s}$); and $4000$ RPM, i.e.,  $419\ \rm rad/s$.

For the representative calculations we take the  3D soft substrate to be composed of agarose hydrogel, and the cortical spheroids and fluid media to be the ones described in \cite{dingle2015three}.
Consequently,
we take the densities of the  3D soft substrate, cortical spheroid,  and  fluid media to be $1640$, $1240$, and $980\ \rm kg/m^3$, respectively. Based on the measurements in \cite{mori2013,normand2000} we take the agarose hydrogel's Lam\'e parameters to be $\lambda=4.28571~\times10^5\ \rm(Pa)$ and $\mu=1.07143\times10^5\ \rm (Pa)$. Based on the measurements in \cite{Boulet2011} we take the shear modulus of the cortical spheroids to be $\mu=1.33 \times10^{3}\ \rm (Pa)$.

We take the values for the geometry parameters to be the ones given in Table \ref{tb:TypicalGeoParams}.

 Some  strain measures from the representative calculations are shown in Figs.~\ref{fig:minstrain}--\ref{fig:strainconponent}, and stress measures in Figs.~\ref{fig:minstress}--\ref{fig:pressure}. The definitions of some of those strain (resp. stress) measures are discussed in \S\ref{sec:StrainCalculations} (resp. \S\ref{sec:StressCalculations}).




\subsection{Strains}
\label{sec:StrainCalculations}
For the strain measure we use the logarithmic strain tensor $\bl{H}_{\bl{\tau}}\ag{\bl{x}}$. The logarithmic strain tensor is defined as
\begin{subequations}
\begin{equation}
\bl{H}_{\bl{\tau}}\ag{\bl{x}}=\ln\bl{V}_{\bl{\tau}}\ag{\bl{x}},
\end{equation}
where
\begin{equation}
\bl{V}_{\bl{\tau}}\ag{\bl{x}}\bl{V}_{\bl{\tau}}\ag{\bl{x}}:=\bl{F}_{\bl{\tau}}\ag{\bl{x}_{\bl{\tau}}^{-1}\ag{\bl{x}}}\pr{\bl{F}_{\bl{\tau}}\ag{\bl{x}_{\bl{\tau}}^{-1}\ag{\bl{x}}}}^{\sf T}.
\end{equation}
\end{subequations}

The co-rotational Cartesian components of $\bl{H}_{\bl{\tau}}\ag{\bl{x}}$ are defined as
\begin{equation}
H_{ij}
\ag{
\usf{x}
}
=
\bl{e}_i\ag{\tau}\cdot
\cbr{
\bl{H}_{\bl{\tau}}\ag{\breve{\bl{x}}\ag{\usf{x},\tau}}\bl{e}_j\ag{\tau}},
\end{equation}
where $\breve{\bl{x}}\ag{\usf{x},\tau}$ is the vector in $\mathbb{E}$ such that $\usf{x}$ is its set of co-rotational Cartesian co-ordinates at the time instance $\tau$.
We denote the matrix $\pr{H_{ij}
\ag{\usf{x}}}_{i,j\in \mathcal{I}}$ as $\usf{H}^{\cbr{\bl{e}_{i}\ag{\tau}}}\ag{\usf{x}}$.

Let
\begin{equation}
\usf{H}^{\cbr{\bl{c}_{i}\ag{\bl{x},\tau}}}\ag{\usf{x}}:=
\usf{R}\ag{\usf{x}}
\cbr{\usf{H}^{\cbr{\bl{e}_{i}\ag{\tau}}}\ag{\usf{x}}}
\usf{R}\ag{\usf{x}}^{\rm \usf{T}}.
\label{eq:lecylindrical}
\end{equation}
We call $\usf{H}^{\cbr{\bl{c}_{i}\ag{\bl{x},\tau}}}\ag{\usf{x}}$, the co-rotational cylindrical components form of $\bl{H}_{\bl{\tau}}\ag{\bl{x}}$.
We denote the $(1,1)$, $(2,2)$, $(3,3)$, and $(1,3)$ components of $\usf{H}^{\cbr{\bl{c}_{i}\ag{\bl{x},\tau}}}\ag{\usf{x}}$, respectively, as $H_{rr}\ag{\usf{x}}$, $H_{\theta\theta}\ag{\usf{x}}$, $H_{zz}\ag{\usf{x}}$, and $H_{rz}\ag{\usf{x}}$.


\begin{figure}[H]
    \centering
        \includegraphics[width=\textwidth]{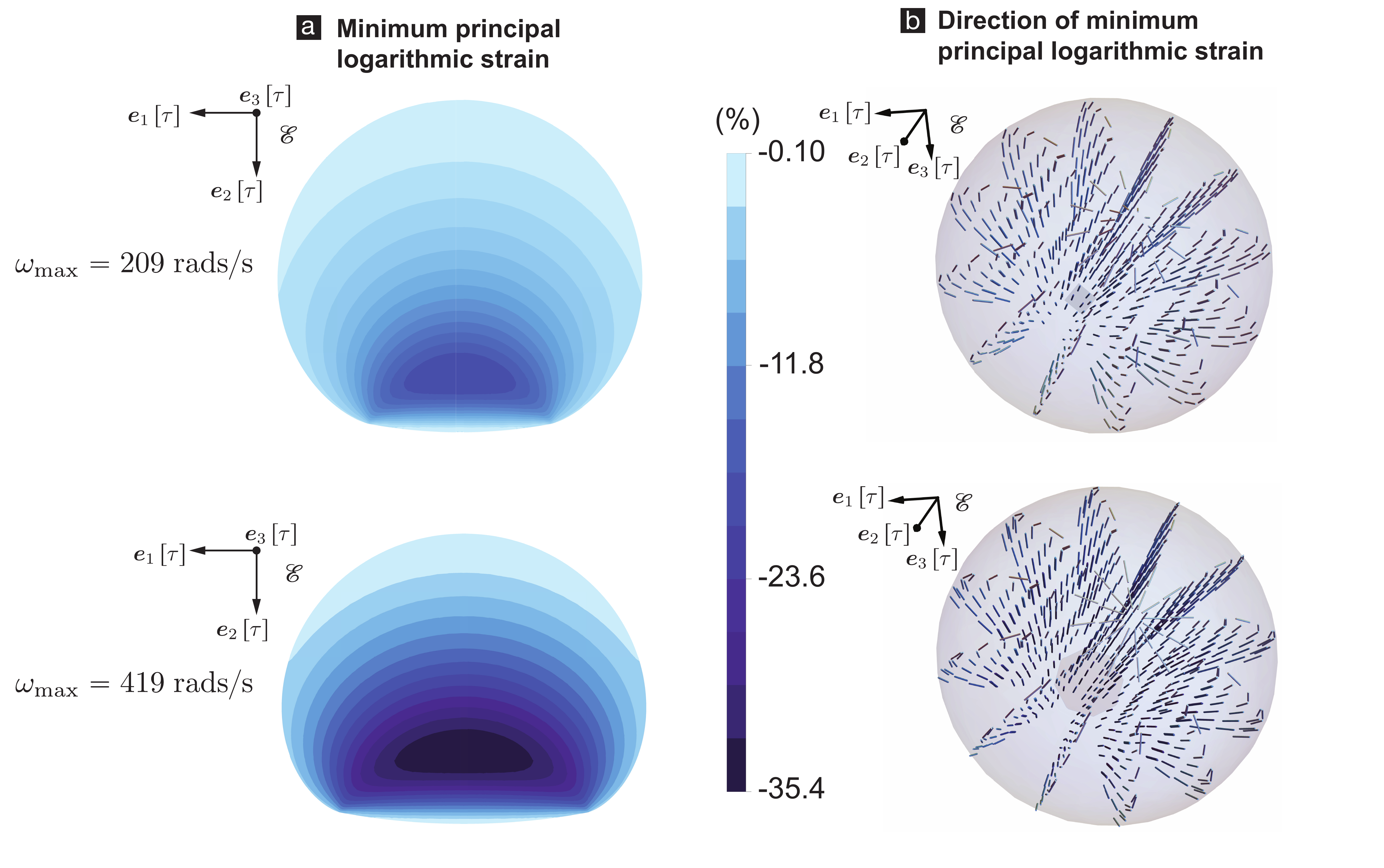}
    \caption{Strains in the spheroid predicted by our theory for the
    representative values of angular velocity, geometry parameters, and material properties
    described in \S\ref{sec:Results} at an arbitrary time instance $\tau$. (a) Contour plots of the minimum principal value of the logarithmic strain tensor. The minimum principal value is the smallest eigenvalue. The logarithmic strain tensor is defined in \S\ref{sec:StrainCalculations}. (b) Each line segment shows a section of the fiber associated with the  eigenvectors that correspond to the  minimum   eigenvalue at the location of its midpoint.
    }
    \label{fig:minstrain}
\end{figure}

\begin{figure}[H]
    \centering
        \includegraphics[width=\textwidth]{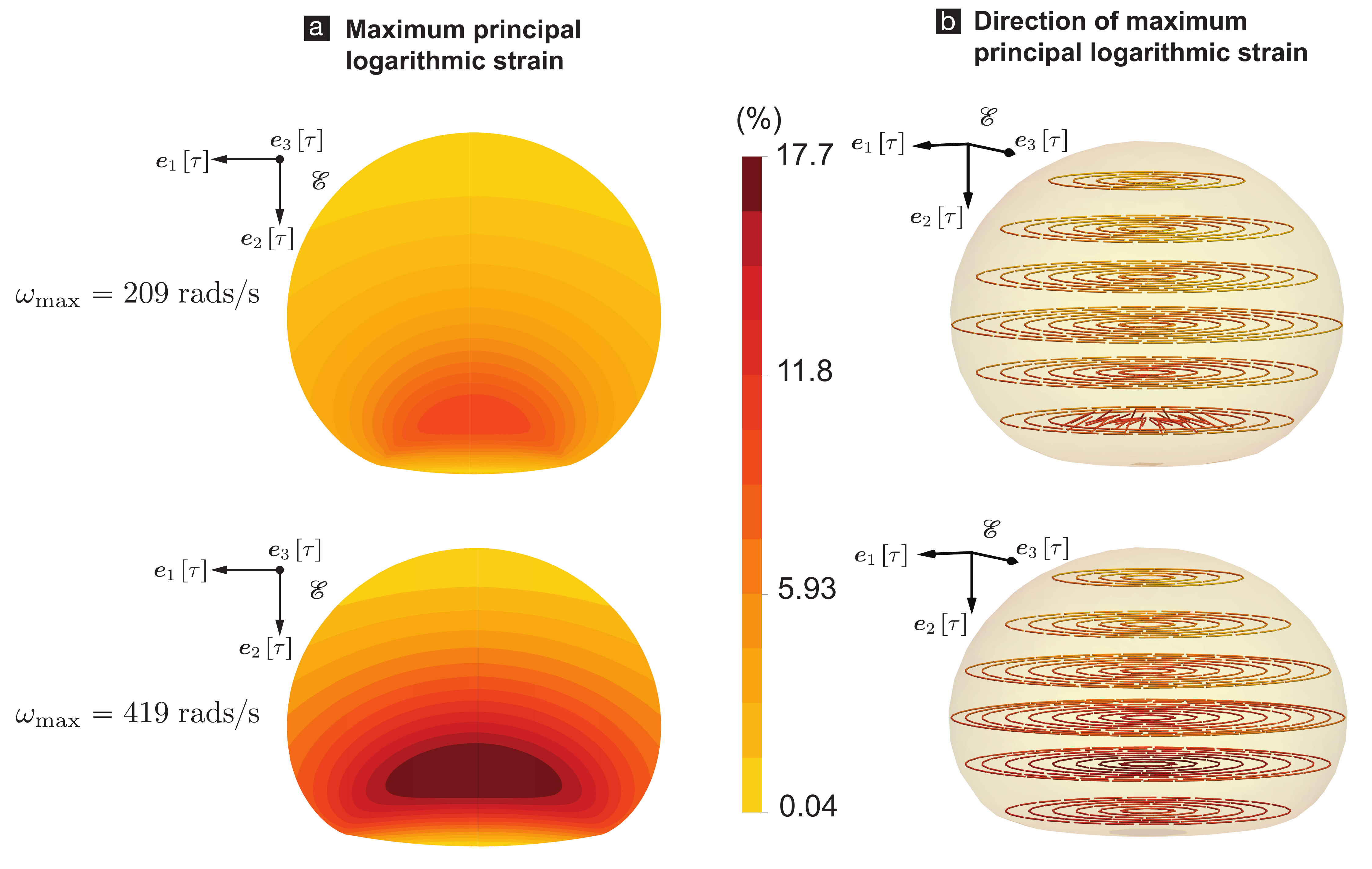}
    \caption{Strains in the spheroid predicted by our theory for the
    representative values of angular velocity, geometry parameters, and material properties
    described in \S\ref{sec:Results} at an arbitrary time instance $\tau$. (a) Contour plots of the maximum principal value of the logarithmic strain tensor. The maximum principal value is the largest eigenvalue. The logarithmic strain tensor is defined in \S\ref{sec:StrainCalculations}. (b) Each line segment shows a section of the fiber associated with the  eigenvectors that correspond to the maximum   eigenvalue at the location of its midpoint.
    }
    \label{fig:maxstrain}
\end{figure}

\begin{figure}[h!]
    \centering
        \includegraphics[width=\textwidth]{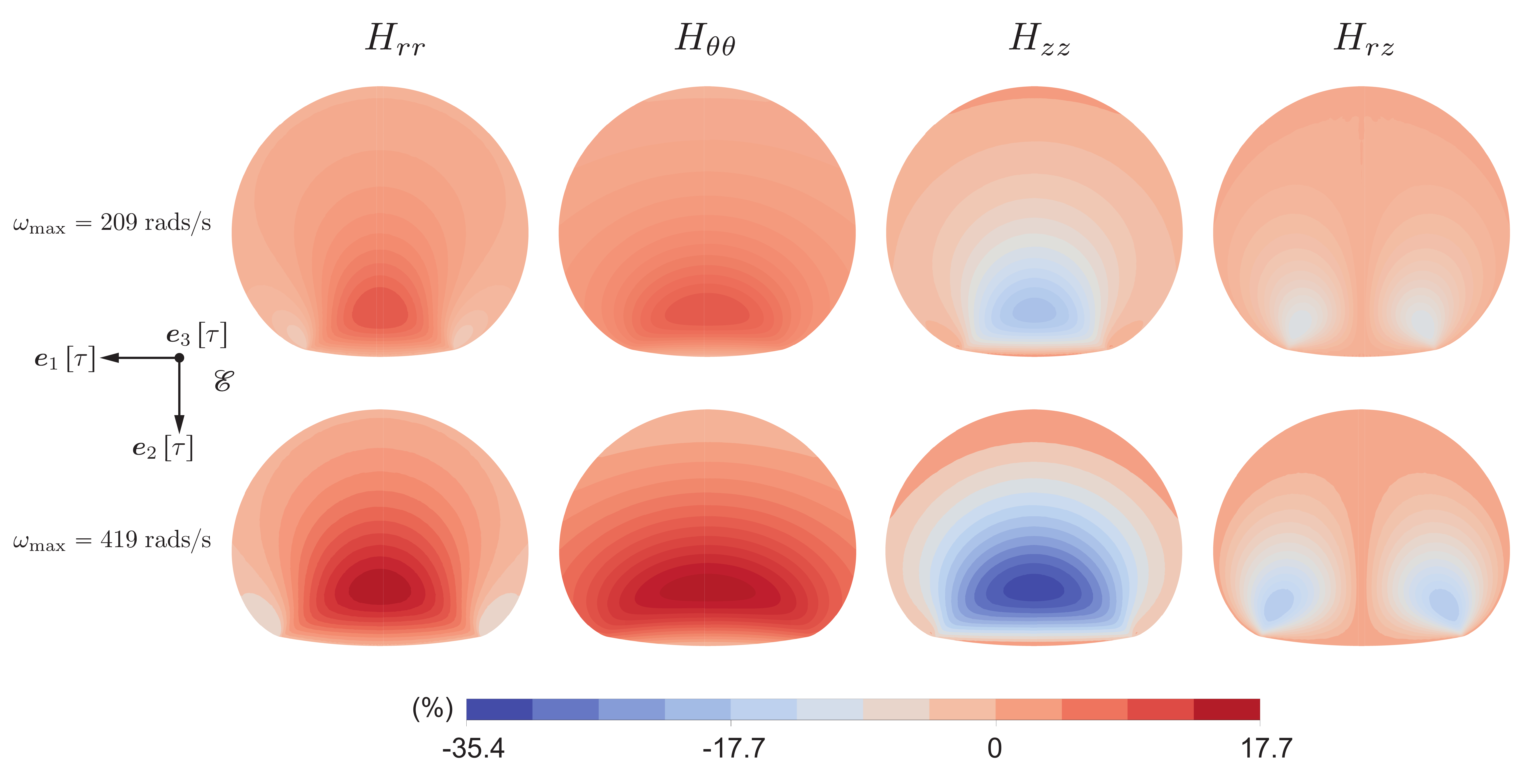}
    \caption{
    Strains in the spheroid predicted by our theory for the
    representative values of angular velocity, geometry parameters, and material properties
    described in \S\ref{sec:Results} at an arbitrary time instance $\tau$.  The logarithmic strain tensor is defined in \S\ref{sec:StrainCalculations}.
    The columns show contour plots of  $H_{rr}$, $H_{\theta\theta}$, $H_{zz}$, and $H_{rz}$, respectively, which are the co-rotational cylindrical components of the logarithmic strain tensor. They are defined in \S\ref{sec:StrainCalculations}. The top row corresponds to the angular velocity $209\ \rm rad/s$, and the bottom row to $419\ \rm rad/s$.
    }
    \label{fig:strainconponent}
\end{figure}

\subsection{Stresses}
\label{sec:StressCalculations}
For the stress measure we use the Cauchy stress tensor.
In \S\ref{sec:StrainCalculations} we defined the co-rotational Cartesian components of the logarithmic strain tensor, $\usf{H}^{\cbr{\bl{e}_{i}\ag{\tau}}}\ag{\usf{x}}$, using the logarithmic strain tensor, $\bl{H}_{\bl{\tau}}\ag{\bl{x}}$, and the co-rotational Cartesian basis $\pr{\bl{e}_i\ag{\tau}}_{i \in \mathcal{I}}$.
In the same way we can define the co-rotational Cartesian components of the Cauchy stress tensor, $\usf{T}^{\cbr{\bl{e}_{i}\ag{\tau}}}\ag{\usf{x}}$, using the Cauchy stress tensor, $\bl{T}_{\bl{\tau}}\ag{\bl{x}}$, and  $\pr{\bl{e}_i\ag{\tau}}_{i \in \mathcal{I}}$. In \S\ref{sec:StrainCalculations} we further defined the co-rotational cylindrical components of the logarithmic strain tensor, $\usf{H}^{\cbr{\bl{c}_{i}\ag{\bl{x},\tau}}}\ag{\usf{x}}$, using $\usf{H}^{\cbr{\bl{e}_{i}\ag{\tau}}}\ag{\usf{x}}$ and the function $\usf{R}\ag{\cdot}$ via \eqref{eq:lecylindrical}.
In the same way we can define the the co-rotational cylindrical components of the Cauchy stress tensor, $\usf{T}^{\cbr{\bl{c}_{i}\ag{\bl{x},\tau}}}\ag{\usf{x}}$,
using
$\usf{T}^{\cbr{\bl{e}_{i}\ag{\tau}}}\ag{\usf{x}}$ and $\usf{R}\ag{\cdot}$.



We denote the $(1,1)$, $(2,2)$, $(3,3)$, and $(1,3)$ components of $\usf{T}^{\cbr{\bl{c}_{i}\ag{\bl{x},\tau}}}\ag{\usf{x}}$, respectively, as $T_{rr}\ag{\usf{x}}$, $T_{\theta\theta}\ag{\usf{x}}$, $T_{zz}\ag{\usf{x}}$, and $T_{rz}\ag{\usf{x}}$.

The pressure at the location $\bl{x}$ at the time instance $\bl{\tau}$ is defined as negative one third the trace of $\bl{T}_{\bl{\tau}}\ag{\bl{x}}$.

\begin{figure}[H]
    \centering
        \includegraphics[width=\textwidth]{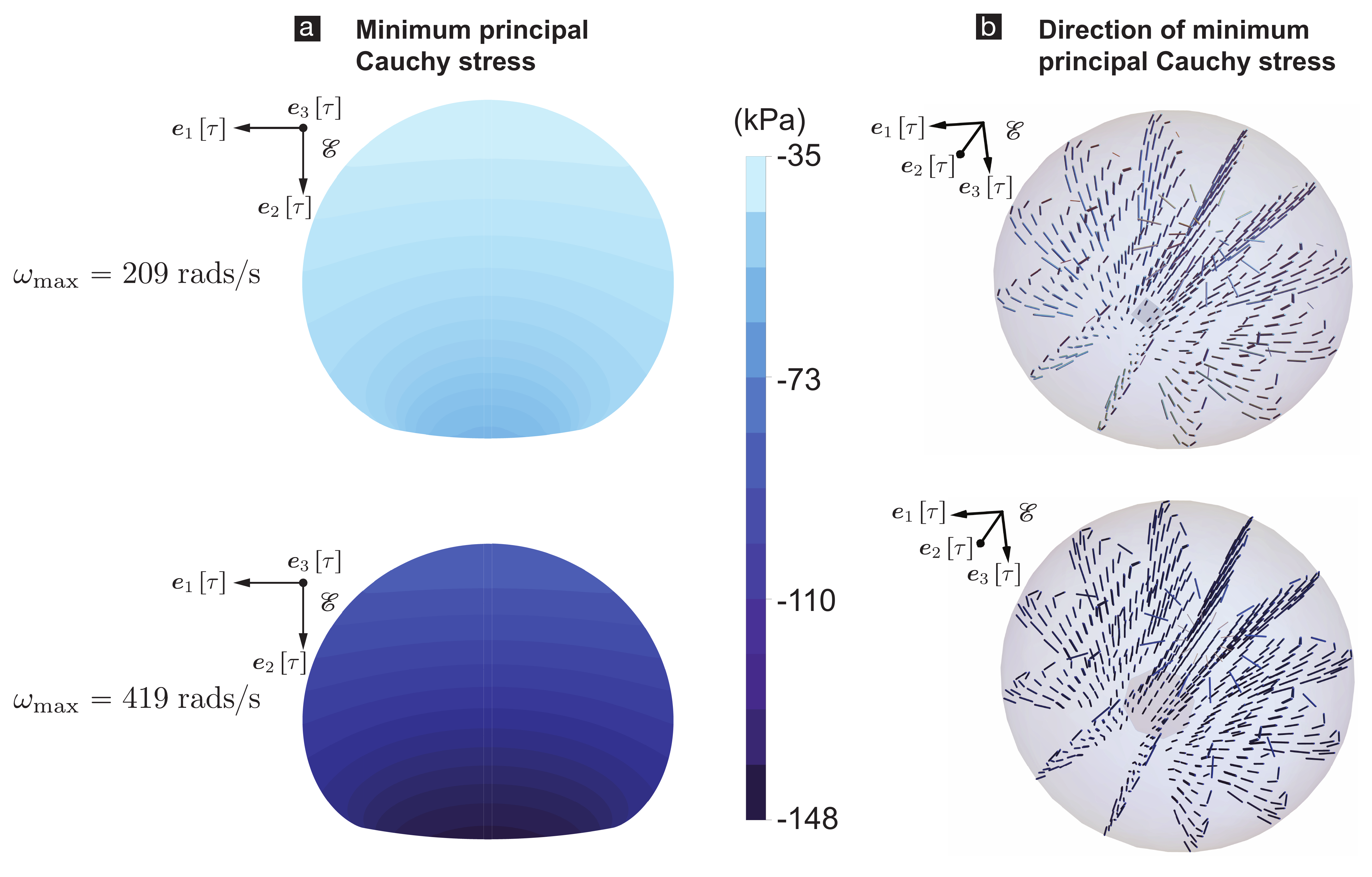}
    \caption{Stresses in the spheroid predicted by our theory for the
    representative values of angular velocity, geometry parameters, and material properties
    described in \S\ref{sec:Results} at an arbitrary time instance $\tau$. (a) Contour plots of the minimum principal value of the Cauchy stress tensor. The minimum principal value is the smallest eigenvalue. (b) Each line segment shows a section of the fiber associated with the  eigenvectors that correspond to the  minimum   eigenvalue at the location of its midpoint.
    }
    \label{fig:minstress}
\end{figure}

\begin{figure}[H]
    \centering
        \includegraphics[width=\textwidth]{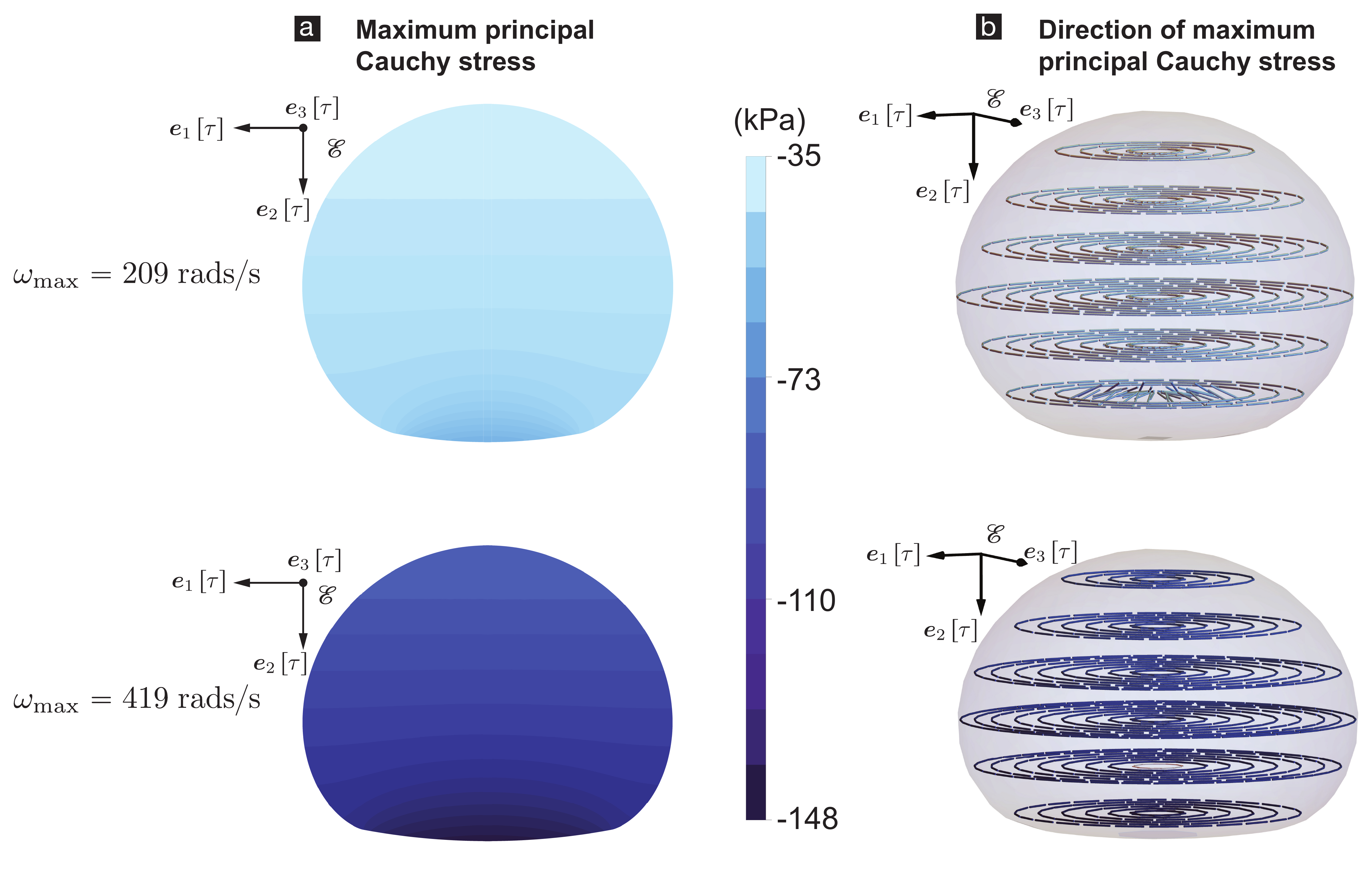}
    \caption{Stresses in the spheroid predicted by our theory for the
    representative values of angular velocity, geometry parameters, and material properties
    described in \S\ref{sec:Results} at an arbitrary time instance $\tau$. (a) Contour plots of the maximum principal value of the Cauchy stress tensor. The maximum principal value is the largest eigenvalue. (b) Each line segment shows a section of the fiber associated with the  eigenvectors that correspond to the  maximum   eigenvalue at the location of its midpoint.
    }
    \label{fig:maxstress}
\end{figure}

\begin{figure}[H]
    \centering
        \includegraphics[width=\textwidth]{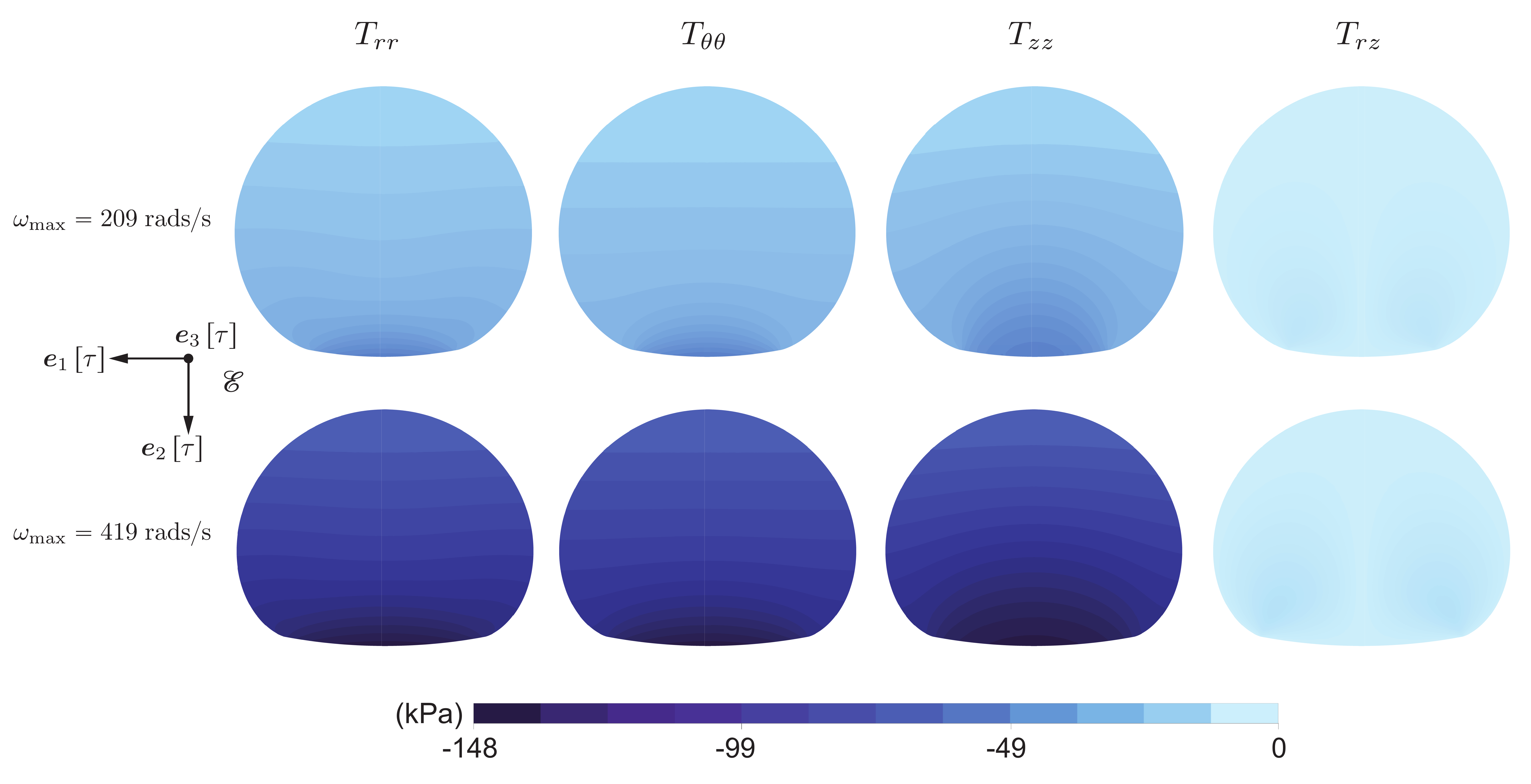}
    \caption{Stresses in the spheroid predicted by our theory for the
    representative values of angular velocity, geometry parameters, and material properties
    described in \S\ref{sec:Results} at an arbitrary time instance $\tau$.
    The columns show contour plots of  $T_{rr}$, $T_{\theta\theta}$, $T_{zz}$, and $T_{rz}$, respectively, which are the co-rotational cylindrical components of the Cauchy stress tensor (see \S\ref{sec:StressCalculations} for details). The top row corresponds to the angular velocity $209\ \rm rad/s$, and the bottom row to $419\ \rm rad/s$.
    }
    \label{fig:stressconponent}
\end{figure}

\begin{figure}[H]
    \centering
        \includegraphics[width=0.8\textwidth]{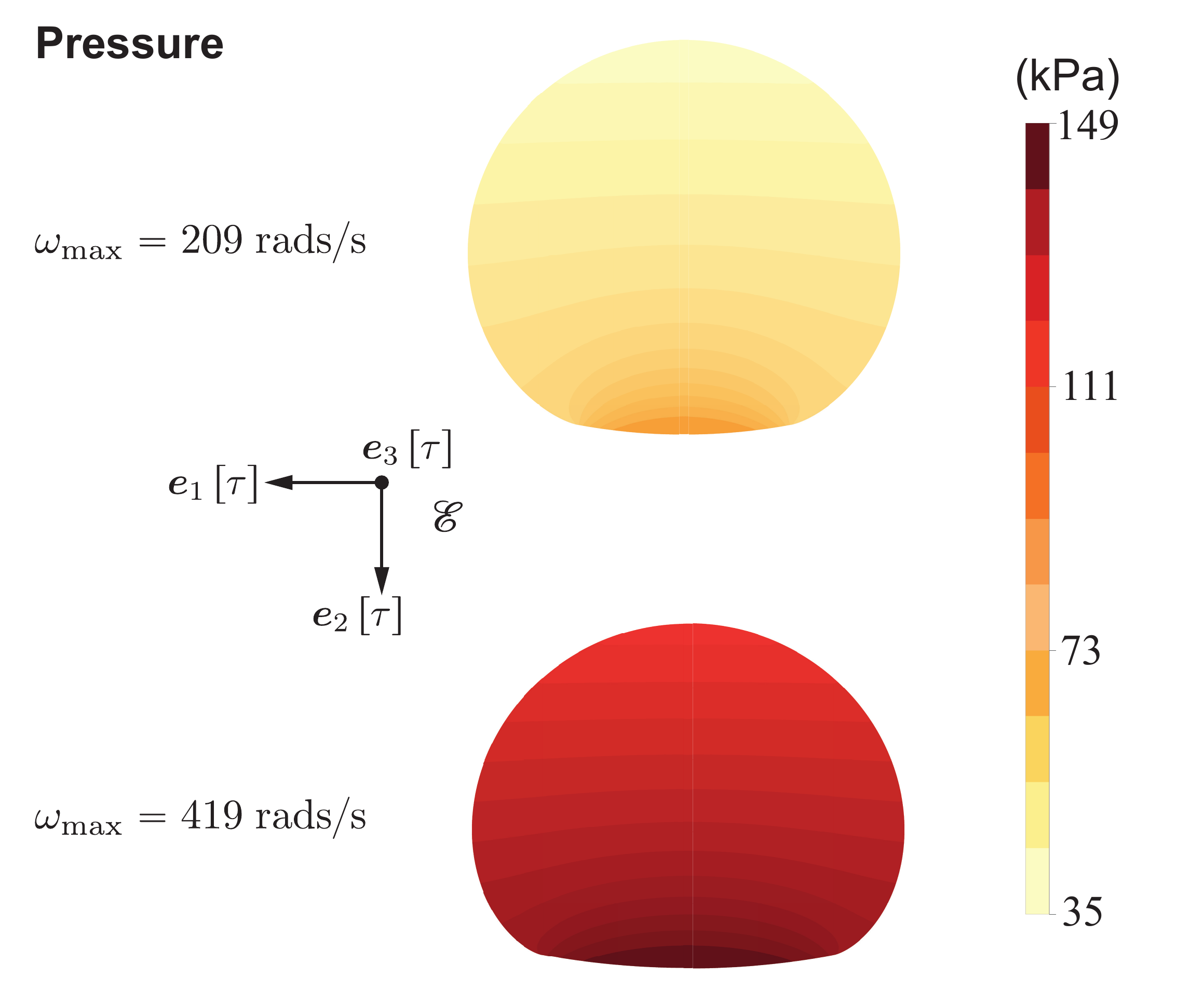}
    \caption{
    Pressures in the spheroid predicted by our theory for the
    representative values of angular velocity, geometry parameters, and material properties
    described in \S\ref{sec:Results} at an arbitrary time instance $\tau$.
    The top plot corresponds to the angular velocity $209\ \rm rad/s$, and the bottom plot to $419\ \rm rad/s$.
    }
    \label{fig:pressure}
\end{figure}

\section{Concluding remarks}
\label{sec:concluding}

\begin{enumerate}

\item In the proposed centrifuge-TBI-model design (Fig.~\ref{fig:centrifuge}), the cortical spheroids  primarily undergo a type of squeezing deformation (Fig.~\ref{fig:centrifuge} (c)). However, with more sophisticated designs for the 3D soft substrate, it is possible to apply other types of deformations to the cortical spheroids.

\item
We have stated earlier (see beginning of \S\ref{sec:assumps}) that we restrict ourselves to the \textit{in vitro} experiment  in which the  centrifuge's   angular velocity is constant as a function of time.
However, based on some preliminary experiments and noting that most micro-tissues are viscoelastic in nature, we  believe that the final results from our theory  will continue to apply with a minor modification even when the angular velocity is not a constant.
Say the angular velocity changes with time as dictated by the function $\tau \mapsto \breve{\omega}_{\rm max}\ag{\tau}$. Then we believe that our results modified by replacing $\omega_{\rm max}$ in them with $\breve{\omega}_{\rm max}\ag{\tau}$ will apply at the time instance $\tau$.
A caveat for our  modified results to apply is that the variation of angular velocity with time be of a moderate character.
That is, at the least, the derivative of $\breve{\omega}\ag{\cdot}$ be well defined and bounded. Since, clearly, our modified results will   not apply when the angular velocity is changed abruptly, i.e., when $\breve{\omega}\ag{\cdot}$ is a step function. (Such a step change in angular velocity will agitate the fluid media in addition to creating other complications.)
Full 3D solution of the Navier-Stokes equations, in the context of the centrifuge-TBI-model, are needed  in order to precisely determine the regime of applicability of our modified results.




\item From a mechanics and engineering perspective, we do not see anything that curtails one from using the idea of applying mechanical loads via centrifugation in other mechanobiology studies.
Especially, \textit{in vitro} traumatic injury studies can be envisioned with other micro-tissues, such as those composed of lung or liver cells.

\item As we mentioned in \S\ref{sec:Introduction}, the mechanical loading, i.e., the force on the cortical spheroid, can be  easily and robustly tuned via the centrifuge's angular velocity and the volume of the fluid media.
As we highlighted in Fig.~\ref{fig:centrifuge} (c), the forces acting on the cortical spheroid consist of the tractions from the fluid media, tractions from the 3D soft substrate, and the body forces due to the rotations.
The fact that the tractions on the cortical spheroid from  the fluid media  depend on the centrifuge's angular velocity and the volume of the fluid media can be seen from    \eqref{eq:pflMat1}.

\end{enumerate}

\section*{Acknowledgments}
The authors gratefully acknowledge support from the Panther Program and the Office of Naval Research (Dr. Timothy Bentley) under grant N000142112044.

\section*{Declaration of Competing Interest}
The authors declare that they have no known competing financial interests or personal relationships that could have appeared to influence the work reported in this paper.

\appendix

\section{Vanishing of the rate of deformation tensor}
\label{sec:D0}
The rate of deformation tensor is defined as
\begin{equation}
\bl{D}_{\bl{\tau}}\ag{\bl{x}}=\frac{1}{2}\pr{\bl{L}_{\bl{\tau}}\ag{\bl{x}}+\bl{L}_{\bl{\tau}}\ag{\bl{x}}^{\sf T}},
\label{eq:D}
\end{equation}
where $\bl{L}_{\bl{\tau}}\ag{\bl{x}}$ is the spatial velocity gradient, defined by
\begin{equation}
\bl{L}_{\bl{\tau}}\ag{\bl{x}}=\left\{\bl{\nabla}_{\bl{x}}\bl{v}_{\bl{\tau}}\right\}\ag{\bl{x}}.
\label{eq:L}
\end{equation}
From \eqref{equ:vel3}, \eqref{eq:L} can be written as\begin{equation}
\bl{L}_{\bl{\tau}}\ag{\bl{x}}=W_{ij}\ag{\tau}\bl{v}_i\ag{\tau}\otimes \bl{e}_j\ag{\tau}.
\label{eq:L1}
\end{equation}
From the  definition of $W_{ij}$ \eqref{eq:WComps}, it follows that $W_{ij}=-W_{ji}$. Then from \eqref{eq:L1} and \eqref{eq:D} we get that
\begin{equation}
    \bl{D}_{\bl{\tau}}\ag{\bl{x}}=\bl{0}.
\end{equation}

\bibliography{refer}

\end{document}